\documentstyle[11pt,aaspp4,psfig]{article}

\begin{document}

\title{Eclipsing Binaries in the OGLE Variable Star Catalog.I.\\
W~UMa-type Systems as Distance and Population Tracers \\
in Baade's Window}

\author{Slavek M. Rucinski\altaffilmark{1}\\
e-mail: {\it rucinski@astro.utoronto.ca\/}}

\affil{81 Longbow Square, Scarborough, Ontario M1W~2W6, Canada}
\altaffiltext{1}{Affiliated with the Department of Astronomy,
University of Toronto and Department of Physics and Astronomy,
York University}

\centerline{\today}

\begin{abstract}
The paper demonstrates that the
W~UMa-type systems have a potential of playing
an important role in studies of the Galactic structure due
to their high spatial frequency of occurrence, ease of detection and
reasonably tight absolute-magnitude (period, color) calibration.

An algorithm, based on Fourier decomposition of light curves, 
permitted to define a sample of 388 contact binaries with well observed
light curves, periods shorter one day and with available $V-I$ colors 
(the R-sample), from among 933 eclipsing binary systems in the OGLE
variable-star catalog for 9 fields of Baade's Window. The sample
of such systems which was visually classified by the OGLE project 
as EW-type binaries
(the O-sample) is by 55\% larger and consists of 604 stars.
The algorithm prevents inclusion of RR~Lyr and SX~Phe stars which 
in visual classification might be mistakenly 
taken for contact binaries with 
periods equal to twice their pulsation periods. 

Determinations of
distances for the contact systems, utilizing the $M_I = M_I(\log P, V-I)$ 
absolute-magnitude 
calibration and the map of reddening and extinction of Stanek
\markcite{sta1} (1996), indicate an approximately 
uniform distribution of contact binaries
almost all the way to the Galactic Bulge (there is a hint of a gap at
6.5 to 8 kpc), with 9 well observed systems identified in the
Bulge. The distances have been 
derived assuming two hypotheses: (1)~extinction extends uniformly all
the way to the Bulge at $d_0 = 8$
kpc or (2)~extinction is truncated at $d_0 = 2$ kpc. Analysis of the
period--color diagram favors the latter hypothesis 
which has been assumed throughout the paper. The uniform distribution
of the contact systems
with distance, implying heights up to $z \le 420 - 450$ pc, as well as
a tendency for their colors to be concentrated in the region 
normally occupied by old Turn-Off-Point stars,
confirm the currently held opinion that contact binary
systems belong to the old stellar population of the Galaxy. 

A first attempt to construct the luminosity function for contact
binaries has been made for samples defined by distances of 2 and 3
kpc. The {\it apparent\/} density of contact systems
is about $(7 - 10) \times 10^{-5}$
systems per pc$^3$, with the main uncertainty coming from the
definitions of the R- and O-samples. If the {\it spatial\/} 
density (corrected for undetected low-inclination systems)
is two times higher than the lower
limit of the above range, and equals 
$1.5 \times 10^{-4}$ systems per pc$^3$, 
then one star among 400 Main Sequence stars
is a contact system; however, 
this number includes M-dwarfs among which contact
binaries do not occur. An independent estimate of 
the {\it apparent\/} 
frequency, relative to nearby dwarfs with colors similar
to those of the contact binaries, obtained in the volume-limited
sense to 2 and 3 kpc, is one contact system per about 250 -- 300 
Main Sequence stars, which agrees well
with the recent estimates for old open clusters and their
background/foreground fields.

\end{abstract}

\keywords{binaries: close --- binaries: eclipsing --- Galaxy: structure
--- gravitational lensing}


\section{INTRODUCTION}

The Optical Gravitational Lensing Experiment (OGLE) is a long term
observing project aimed at detecting invisible matter in our Galaxy
by detection of gravitational microlensing events in the direction of
the Galactic Bulge. It has been giving a rich spectrum of by-product
results, among them excellent color--magnitude diagrams (Udalski et
al. \markcite{uda1}
1993) which have provided interesting insight into the structure
of the Galactic Disk (Paczy\'nski et al. \markcite{pa1}
1994a = P94), non-axiality of
the Bulge (Stanek et al. \markcite{sta2}
1994, Paczy\'nski et al. \markcite{pa2} 1994b), 
and maps of the interstellar extinction in Baade's Window (Stanek 
\markcite{sta1} 1996 = S96).

The first utilization of another advantage of this
dataset, of its systematic temporal coverage permitting detection of very
large numbers of periodic variable stars, was presented by 
Rucinski (\markcite{ruc7} 1995a = BWC) 
in a preliminary analysis of 77 contact binaries, from
the first instalment of the periodic variable-star catalog for 
the central Baade's Window field (Udalski et al. \markcite{uda2} 1994).
Since then, the OGLE Project released the Catalog of periodic variable stars
in the remaining 8 fields of Baade's Window (Udalski et al.
\markcite{uda3} \markcite{uda4} 1995a, 1995b), with data now available 
for the whole $40' \times 40'$ field. This
database will be from now on called the Catalog.

The present paper is basically a full version of the BWC study
for whole Baade's Window
and addresses the same issue, which is the utility
of contact binaries (or W~UMa-type or EW-type systems, as we will call
them below) as distance tracers in Galactic structure
studies.  From among all 933 eclipsing systems, 676 
systems with periods shorter than one day were first selected, 
but this number was further substantially reduced to 388 
by acceptance as a contact binary
through a strict criterion based on the Fourier decomposition of the
light curves (Section \ref{alg})
and availability of $V-I$ colors.
 This selection was entirely independent of the original
classification as EW-type systems in the Catalog. We note, that among 
662 EW systems in the Catalog, 604 systems have $V-I$ colors and
periods shorter than one day, so that our (impersonal)
selection of 388 systems imposed more severe restrictions 
on the acceptance as a contact binary than the OGLE classification.

The present paper utilizes the fact that sizes of contact binaries can
be estimated from their periods so that -- with color information --
absolute magnitudes can be derived permitting usage of these systems
as distance tracers (BWC, Rucinski \markcite{ruc5} 
1994a = R94, Rucinski \markcite{ruc9} 1996 = R96).
Although functional dependencies of dynamical time scales are
underlying absolute-magnitude calibrations for both, the RR~Lyrae-type
stars and the contact binaries, the latter are probably
less good standard candles than RR~Lyrae stars, mostly because of the
relatively strong color dependence (but infra-red observations might
help here). Depending on the combination of the color and period, the
absolute magnitudes for contact binaries
 can be estimated to  0.2 -- 0.5 mag. The spread for
the RR~Lyrae stars of a given metallicity is some 2 -- 4 times smaller
(cf.\ recent results for M5 by Reid \markcite{rei1} 
1996), although for large,
inhomogeneous samples it might be substantially larger. We note that the
small spread in luminosities of the RR~Lyrae stars is simply due to
their occurrence in a very small region of the stellar parameter
space. In this respect, the contact binaries have a great
advantage: They occur with high frequency (as we show later,
in the solar neighborhood, they are some 24,000 more common than
RR~Lyrae stars!) and, by being slightly evolved,
they trace the most common Main Sequence population to greater
depths than do ordinary single stars.
Also, apparently their luminosities
are relatively insensitive to $[Fe/H]$ (Rucinski \markcite{ruc8}
1995b = R95).
Besides, the RR~Lyr stars are present only in very old stellar systems,
older than about 10 Gyr, whereas contact binaries seem to
start appearing in
open clusters with ages around 1 Gyr (Ka\l u\.zny \& Rucinski 
\markcite{kal3} 1993, Rucinski \& Ka\l u\.zny \markcite{ruc10}
1994). The luminosity calibrations for
the contact binaries do have weaknesses related to occasional
occurrence of spots on some systems and
lack of good calibrators over the whole period--color sequences
(Rucinski \markcite{ruc9}
1996), but -- hopefully -- the large numbers of these 
systems in the micro-lens databases might help compensate for these
deficiencies or reveal trends which could be explored in more
in-depth studies.

Origins and evolution of contact binary stars are still 
poorly understood although their low angular momenta, presence in old
open and globular clusters (Ka\l u\.zny \&
Rucinski \markcite{kal3} 1993, Rucinski \& Ka\l u\.zny 
\markcite{ruc10} 1994)
and space motions (Guinan \& Bradstreet \markcite{gui} 1988) suggest
formation by a relatively slow process of
coalescence of close, detached binaries through magnetic-wind induced
loss of angular momentum. Advanced evolution is particularly strongly
 suggested by occurrence of contact binaries among
Blue Straggler stars in old open and globular clusters (Mateo 
\markcite{mat1} 1993, \markcite{mat2} 
1996). There have been several general
reviews on the subject of contact binaries, the most recent ones by
Rucinski \markcite{ruc1} (1993a) and Eggleton \markcite{eggl}
(1996). Of interest for this work are results of 
recent systematic searches in stellar clusters which have led to an
upward revision of the {\em apparent\/} frequency, 
both in clusters and in the Galactic field, to about one contact
system per 280 stars (R94), with the implied {\em spatial\/} 
frequency (corrected for undetected systems at
low orbital inclinations) of about 1/140, in agreement with entirely
independent estimates of Mateo which were based on numbers of stellar
X-ray sources (Hut et al. \markcite{hut} 
1992).  The contact systems are 
also recognized as extremely important for
dynamical evolution of globular clusters (Hut et al. \markcite{hut}
1992).

There exists an abundance of light curve data for many bright contact
binaries in the sky, but the material is  highly inhomogeneous 
in terms of accuracy as well as sampling of the parameter
space. Perhaps the most troublesome is lack of standard color data
even for quite bright W~UMa-type systems.
The new micro-lensing samples contain 
more uniform data than presently available for field systems, the
reason being that the latter in their majority have
come from very non-systematic 
searches of the sky, with equally non-systematic follow-up studies. 
The General Catalogue of Variable
Stars lists 562 contact systems, but for only 510 of them are 
the  periods reliably known, whereas moderately good photometric 
data exist for some 130 systems, but again only for half of 
that in standard photometric systems. Thus, although the amount of
information for each system in the OGLE Catalogue is limited,
its quality is uniform or at least controllable, and can be easily
subject to statistical scrutiny. 

Section~\ref{ecl} of 
the present paper contains a brief assessment of the database for
eclipsing systems in the Catalogue. Section~\ref{alg}
 describes an algorithm
used to select a sample of contact binaries used as a primary dataset
in the paper (the R-sample). Section~\ref{dist}
 discusses estimates of observed
and absolute magnitudes of the systems and then of their
distances. Section~\ref{pop}
 discusses implications of the present results for
our understanding of the contact systems 
and Section~\ref{sum} gives a summary of the paper.


\section{ECLIPSING BINARY SAMPLE}
\label{ecl}

The sample of eclipsing binaries in the OGLE Catalogue (Udalski et
al. \markcite{uda2} \markcite{uda3} \markcite{uda4}
1994, 1995a, 1995b) contains 933 entries. 
The data listed in the Catalogue consist of: the $\alpha_{2000}$ and
$\delta_{2000}$ coordinates, 
$I_{max}$, $(V-I)_{max}$, $\Delta I$, 
the period and zero epoch, and the light
curve (typically 100 -- 190 points) in $I$.
The sample contains systems with $14 < I < 18 $, but selection effects
against discovery might be expected at fainter magnitudes
due to progressively stronger blending of images. A
histogram of magnitudes at maximum light, $I_{max}$, shown in Figure~1,
indicates an increase in numbers of detected eclipsing systems 
with decrease in brightness which is
surprisingly consistent with the uniform-density, empty-space law of
``4-times-per-magnitude'' increase. 
A clear deficiency relative to this law
shows up for $I_{max} > 17.0$ and an abrupt cut-off takes place at 
$I_{max} = 17.9$, close to the stated limit of the catalog at $I =
18$. Possibility of selection effects at fainter
magnitudes led us to consider in Section~\ref{dist} the 
numbers of contact systems in relation to the
total numbers stars in Baade's Window rather than in absolute sense.

The histogram of amplitudes shown in Figure~2 indicates that selection
effects against discovery
might be present for amplitudes $\Delta I$ smaller than about 0.3,
although this distribution cannot be obviously separated from the
intrinsic distribution of the variability amplitudes. 



The random errors in the data were estimated by evaluating spread in
light curve points at phases close to phase 0.25 for all eclipsing
systems in the Catalog. This was done in a
re-normalization step, while
fitting the light curves by Fourier cosine series, as described in the
next section. The distribution of errors with brightness is shown in
Figure~3 where it is separated into four magnitude bins.  From
inspection of this figure, we conclude that typical random errors in
$I$ (per observation) were about 0.02 for $I < 17$ and increased
somewhat to about 0.03 -- 0.04 for $I > 17$.



\section{AUTOMATIC CLASSIFICATION OF CONTACT BINARIES}
\label{alg}

The OGLE Project, when releasing the variable-star Catalog attempted a
visual classification of variability on the basis of morphology of
light curves. While separation of contact binaries from other
eclipsing systems in most cases does not present problems,
it would be useful to have an automatic classifier
which would be impersonal, easy to repeat and -- if need occurs 
-- equally easy to modify. We would also like to have a simple way of
removing poorly observed systems as these might be stars of other
types (spotted, pulsating) or might have poor determinations of
periods. Such a classifier has been constructed here on the basis of
the Fourier decomposition of light curves. As was discussed in
Rucinski \markcite{ruc4}  (1993b = R93), light curves of contact 
systems are very simple in their shape and
usually only two cosine coefficients are needed to adequately describe
a light curve. Although this revealed a basic difficulty of extracting
several geometric elements from light curves of such a low information
content, this is a convenient circumstance for large databases of
variables observed with moderate accuracy, as in the case of the OGLE
Catalog. 

As was described in R93, comparison of the cosine coefficients $a_2$
and $a_4$ permits a crude estimate of the degree of contact which is
weakly dependent on the inclination and mass-ratio of the system. This
property was noted already in Rucinski \markcite{ruc2}
(1973) and explained through
different sensitivity of $a_2$ and $a_4$ to the global distortion of
the contact structure and to the more localized eclipse
effects. However, it remained unclear whether the  Fourier
coefficients would retain usefulness for detached systems as no
single database existed which could be used to compare -- in large
numbers -- light curves of detached and contact eclipsing system. The
usefulness of the approach has been proven positively here 
by application of the Fourier coefficients to
the whole sample of the OGLE eclipsing systems.

The light curves of all 933 systems were analyzed exactly following
the prescriptions in R93 and the results in the $(a_2, a_4)$ plane are
shown in Figure~4. Note that the light curves for the Fourier analysis
were expressed in light units and each time a new normalization was
done for phases around 0.25 (at least 10 observations). 
This normalization
permitted to estimate spread of observations at these phases and gave
information of measurement errors, as shown in Figure~3 in previous
section. 

In Figure~4, the systems have been marked differently following
the OGLE classification: EW or contact binaries of the W~UMa type
(filled circles), EB or systems showing Beta Lyrae light 
curves with unequal minima (open circles), and EA and E 
detached binaries and binaries of uncertain type (crosses). 
A theoretical envelope based on the results in R93 for the inner
(marginal) contact is shown as a continuous curve (see below).
The EW-type systems are indeed
confined below the theoretical limit for marginal contact (with a
well-known tendency for weak contact),
while most of the systems classified as detached eclipsing (EA, E and
other uncertain, 250 systems in total) are
above this limit. The relatively small number of systems classified as
Beta Lyrae-type (EB, 21 systems) basically follow the division line
between the detached and contact groups. In
what follows we simply disregarded this group and included those
systems in the contact-binary sample which passed the Fourier
filter. We note that some genuine contact systems show unequal minima 
indicating poor thermal contact. This matter will remain open until
the $V-I$ color curves become available, as then the correlated
behaviors of the $a_1$ coefficients for the light and color curves
could be used as a detection criterion for EB systems. 

Once the color-curves become available, they will provide useful criteria 
to guard against inclusion of pulsating stars with 
periods equal to twice their pulsation periods.
At present, we rely entirely on the Fourier coefficients. In
particular, negativity of $a_4$ prevents inclusion of the RR~Lyrae
stars, although the RRc-subclass might occasionally create problems for
less accurate light curves. As an illustration, we show in Figure~4 the
$(a_2, a_4)$ coefficient pairs for the pulsating stars in one (central)
field of the OGLE project. 


The envelope for the inner contact in Figure~4
can be approximated very well by
a following simple relation 
(cf. Figure~6 in R93): $a_4 = a_2(0.125-a_2)$, where both
coefficients are negative. In this paper, 
position relative to the inner-contact curve is the
main criterion for classifying a star
as of the EW-type  (below the curve) or other eclipsing  
(above the curve). This criterion was supplemented by  one
measuring the over-all quality of fit by a simplified
light curve consisting only
of 5 Fourier terms: $l(\theta) = \sum_{i=0}^4 a_i \cos 2\pi i\theta$. 
The fit was usually excellent for contact
binaries, whereas for detached binaries it showed large systematic
deviations which made the formal errors  of the
coefficients $a_i$ very large and thus basically useless in the
analysis. Typical fits are shown in Figure~5 for the first
EW and EA systems of the OGLE sample, V4 and V19, in the field
BWC\footnote{Hereinafter, for identification of individual
systems, we will be using designations of the following
 type: \#$n.ddd$, where
$n$ is the field number (for the Central field $n=0$), and $ddd$ is
the variable number in the field.}. 


The $(a_2, a_4)$ diagrams for different ranges of the
mean standard errors of the 5-term fits, in light units, are shown in
Figure~6. Whereas 525 systems with periods shorter than one day
passed the $(a_2, a_4)$ filter, the requirement that the overall
Fourier fit be better than 0.04 eliminated most of the detached and
many poorly-observed contact systems leaving 404 objects. 
Among these, 388 systems have $V-I$ colors at maximum light. 
The sample of contact systems selected in this way will be, from now
on, called the ``restricted'' sample or the R-sample. We feel that
although this sample might have been somewhat conservatively selected, it
does contain only genuine and well-observed
contact systems. Most of analyses were done
for this sample. The sample of EW systems
selected by the OGLE project, limited to periods shorter than one day
and with available colors consist of 604 systems and, 
from now on, will be called the O-sample. Spot checks of those
systems which belong to the O-sample, but were rejected for the
R-sample fully confirm that, indeed, 
these cases should not be considered,
either due to large measurement errors (most
probably due to seeing-dependent
blending of images) or to light curves with partial
phase coverage.


A more extensive discussion of information contained in
light curves of the eclipsing binaries
in the Catalog is planned for a separate paper of this series.

\section{CONTACT BINARY SYSTEMS AS DISTANCE TRACERS}
\label{dist}

\subsection{The Observed Color--Magnitude Diagram}
\label{cmd}

Several elements are needed to derive distances to the contact
systems. In addition to the observed maximum
magnitudes and colors, $I$ and
$V-I$, one needs values of reddening and extinction as well as
absolute-magnitudes derived from a
calibration utilizing colors and periods. 
These details will be described in the subsequent
subsections. However, much information is already 
contained in the observed
color--magnitude diagram (CMD) for the contact system which is shown
in Figure~7. This diagram shows systems of the R-sample,
together with a schematic representation of the distribution for
normal stars.

Position of one blue system in Figure 7 at $V-I = 0.75$ and $I = 17.4$,
\#6.121 should be noted. Although its light
curve seems to be well defined, a closer examination of the data
by Dr.~Andrzej~Udalski revealed that its color might be affected by image
blending. Another blue system in the O-sample, \#4.177 with $V-I
= 0.67$, has been rejected by the Fourier filter because $a_4$ for it
was positive. However, its color seems to be genuinely blue.
Both systems fall  outside the range where our absolute-magnitude
calibration is strictly applicable so that very large estimates of
their distances might be entirely incorrect.

Besides those two blue systems, we also see unusually red systems or
systems with very short orbital periods. In the R-sample, systems
\#0.160, \#3.053, \#5.143, \#6.123, \#7.112, \#8.072 
have $V-I > 2.0$. Such very red colors are unusual for contact
binaries. At this point, we might only hope that these are not
indications of a systematic deficiency in the color transformations in
the OGLE data at the red end of the color distribution. 
Turning to very short periods: 
The system \#4.040 has the orbital period of 0.228 day with
$V-I=1.69$, whereas the system \#3.038, with the orbital period of only
0.198 day and $V-I=2.45$\footnote{This extremely 
interesting system 
was pointed to the author by Andrzej Udalski before start of this
study.} must remain in the O-sample, as it has not passed
(marginally) the Fourier filter with $a_2=-0.202$ and
$a_4=-0.087$; these coefficients suggest a detached, but 
very close system of strongly distorted, late-type components. 
These special cases will be 
a subject of another paper of this series. Here, we concentrate on
normal systems and limit ourselves to time-independent quantities,
temporary disregarding outliers in the sample.


A schematic outline of the location of the
majority of stars in the OGLE fields in the color--magnitude diagram
is shown in Figure~7 by a shaded area. 
Spatial distributions of these normal, predominantly single stars, were 
analyzed in P94 in terms of the Galactic 
structure in the direction to the Bulge. The slanted sequence of
relatively blue stars was interpreted there as Main-Sequence stars of
a young population, but this interpretation has been recently
replaced (Kiraga, Paczy\'nski \& Stanek \markcite{kir} 1996; 
see also Bertelli et al. \markcite{ber} 1995 and 
Ng et al. \markcite{ng2} 1996) 
by postulating that the sequence is formed by old, Turn-Off-Point 
stars, progressively reddened with distance. 
The contact binaries scatter around this
sequence in Figure 7, so that any inferences that we
could establish for the contact binaries should possibly also relate
to these stars as well.


To visualize the relation of numbers of the W~UMa systems to normal
stars, the next Figure~8 shows sections of equal color 
vertically through the CMD in Figure~7. These sections were defined as 
bands 0.1-magnitude wide in $V-I$
in this part of the CMD where we see most of the contact binaries, i.e.
within $1.0 < V-I < 1.4$ and $16.0 < I < 17.9$. The sample of normal
stars in Figures 7 and 8 
is the one used by Udalski et al. \markcite{uda1}
 (1993) and P94 in their discussions
of the color--magnitude diagrams. These CMD-stars were selected for
best photometry and therefore represent a sample of different stars
than those which were studied for periodic variability. 
An arbitrary normalization factor of 100, 
used in Figure~8, was apparently approximately a correct one to
show both, the CMD stars and the contact binaries, in one figure. In
fact, the plots directly show that the frequency of
occurrence of contact binaries in our sample is
similar to, but exceeds, the apparent frequency for old open
 clusters and for Milky Way background/foreground, estimated from the
recent CCD surveys (R94) at about one W~UMa system per 280 stars. 
Although, for the first time, we have here 
a luxury of being able to obtain the apparent frequency in
color--magnitude bins, and not just by relating numbers of detected
W~UMa-type systems to numbers of observed stars, we should remember
that (1)~the sample of the CMD-stars is biased toward good photometric
quality with many faint stars (estimated here at 45\%)
missing from the statistics,
and (2)~contact binaries are intrinsically brighter than
normal stars, so that the samples are not co-spatial. 
In spite of these qualifications, 
the {\em shapes\/} of the distributions in
Figure~8 are the same to the accuracy of statistical uncertainties
(the reduced $\chi^2 \simeq 0.4 - 1.4$) when the distributions are 
simply normalized to the total number of systems in each color band.
In the four 0.1-mag wide
color bands vertically across the color--magnitude
diagram in Figure~7, from blue to red  we have 13, 33, 81 and 97 contact
binaries of the R-sample, which are out-numbered by CMD-stars by factors of
$153 \pm 42$, $87 \pm 15$, $51 \pm 6$ and $73 \pm 7$,
where the uncertainties are given by Poisson errors. 

We will return to the matter of relative frequency in Sections
\ref{freq} and \ref{local}, in discussions of a
volume-limited sample, after presenting determinations of
absolute magnitudes and distances. To  determine these latter
quantities, an iterative scheme was applied which involved several
components. These will be discussed in the next sub-sections.

\subsection{Absolute Magnitude Calibration}
\label{cal}

The sample of standard W~UMa-type systems which was used to derive
the $M_I = M_I (\log P,\,V-I)$ calibration was
basically the same as in the previous related papers (BWC, R94, R95),
but the data have been 
somewhat improved. The final calibration (Rucinski 1996, 
unpublished)
is only slightly different from the one which was used in
BWC. It is based on 19 systems: 3 with known trigonometric parallaxes,
12 in high-latitude open clusters and 4 in visual binaries with
fainter companions which can be placed on the Main Sequence. Relative
to BWC, the major improvements are in that (1)~three systems in M67 and
one in Praesepe have now measured $V-I$ colors which confirm those
previously estimated from $B-V$ colors, (2)~new observations of the
CT~Eri ($P=0.634$ day) in a visual binary permit extension to slightly
longer periods 
than before, (3)~systems in NGC~188 with colors transformed from $B-V$ are
de-emphasized as some important systems there show unexplained
deviations: ER~Cep and ES~Cep
are too bright by $\simeq 0.5 - 0.8$ mag, and
the important evolved system V5 is too faint by $\simeq 0.7$ mag, probably
reflecting poor contact (its relatively red 
color should be even {\em redder\/} in good contact). 
According to the most recent proper-motion study by Dinescu et
al. \markcite{din} (1996), ES~Cep is not a member of NGC~188. 
At present, the calibration extends over the following ranges:
$0.27 < P < 0.63$ day, $0.38 < (V-I)_0 < 1.21$ and $1.8 < M_I <
5.0$.  Applicability of the calibration
beyond the above ranges is unknown at this time. Because of this limitation,
figures subsequent to Figure~10 distinguish (by
filled circles) those cases where, after de-reddening,  systems fall
within the strict ranges in all three quantities.

The equations giving the absolute magnitudes and the coefficients are:
$$ M_I = b_{P(VI)}\,\log\,P + b_{VI}\,(V-I)_0 + b_{0(VI)}$$
\noindent
with  $b_{P(VI)}  =  -4.6 \pm 2.0$, $b_{VI} =  +2.3 \pm 1.3$
and $b_{0(VI)}  =  -0.2 \pm 1.0$, and $\sigma = 0.30$ (per star). 
The bootstrap sampling as in R94 
permitted another set of estimates, in terms
of median and $\pm 1$-$\sigma$ ranges for
the distinctly non-Gaussian distributions of the coefficients:
$b_{P(VI)} = -4.4^{+1.3}_{-1.6}$, $b_{VI} = +2.3^{+0.9}_{-0.6}$
and $b_{0(VI)} = -0.2^{+0.2}_{-0.3}$. The strong correlations
between the coefficients result in smaller uncertainties
in $M_I$ than these large errors in the coefficients would imply.
Monte-Carlo simulations show that within these ranges,
predictions on $M_I$ should be good to 1-$\sigma$ level of about
$\pm 0.2$ and  2-$\sigma$ level of about
$\pm 0.5$ mag, although systematic errors
are still possible, mostly because of spots on some of the
late-type calibrating systems. Comparison with the previous
calibrations (BWC, R95) indicates 
that the period coefficient is slightly larger in the absolute sense
than before and that the color coefficient is smaller. Although this makes
little difference for the final values of $M_I$ (since the terms are
correlated), the new calibration is therefore
somewhat less sensitive to the
uncertainties in the reddening than the previous versions. We note
also that the $(V-I)$-based calibration is relatively insensitive to
variations in $[Fe/H]$ (R95). The lines
of constant $M_I$ are over-plotted in the
period--color diagram in Figure~10 (Section \ref{pc}).


\subsection{Maximum Reddening and Extinction}
\label{red}

To find individual values of the reddening
($E_{V-I}$), extinction ($A_I$), distance ($d$) and absolute
magnitude ($M_I$), an iterative procedure was used. This procedure
will be described in Sec.\ref{det}.
It utilized maps of extinction $A_V$ and reddening $E_{V-I}$
in Baade's Window which have been  recently published by Stanek 
\markcite{sta1} (1996; see also Wo\'zniak 
\& Stanek \markcite{woz} 1996).
This excellent tool permits estimates of 
{\em maximum\/} values of both quantities in this complex region of 
strongly variable and patchy interstellar absorption. Extinction in
the $I$-band was found from: $A_I^{max} = A_V^{max} - E_{V-I}^{max}$.
Both, $E_{V-I}^{max}$ and $A_I^{max}$, are indeed
the maximum values as they were estimated in S96
from the Red Clump stars in the Bulge at about 8 kpc: We do not know
the actual values of $A_V$ and $E_{V-I}$
for locations of our systems in space except that they
should fall somewhere between zero and these largest values at some
distance $d_0$. This distance would measure the effective thickness of the
absorption layer and, by definition,
should be smaller than the distance to the Bulge, here set at 8 kpc.

Although, following S96, independent values of reddening and
extinction were used in most analyses, for some simplified estimates
that will follow, $A_V = 2.5\, E_{V-I}$ was sometimes assumed. 
An approximate
average for the central portion of Baade's Window is $A_V \simeq
1.5$ (Paczy\'nski et al.\markcite{pa1} 
1994a), hence $E_{V-I} \simeq 0.6$ and $A_I
\simeq 0.9$. These median values are a bit smaller than the modal ones
of about 0.65 and 1.0, respectively.
 However, as demonstrated in S96, the range in
extinction over the whole area is much wider, $1.2 < A_V < 2.8$, so
that allowance for positional dependence should substantially
improve the quality of the data over an assumption of an average
extinction. 

The input data for each star were: the position ($\alpha$,
$\delta$), the observed magnitude ($I$), 
the observed color ($V-I$) 
and the orbital period $P$. Two independent 
quantities were taken from the S96 maps: the maximum reddening
$E_{V-I}^{max}$ and the maximum extinction $A_V^{max}$.
Both have been derived by two-dimensional interpolation in the
extinction maps for all systems of the sample.
For 135 cases of stars outside the maps, from among 933 of the OGLE
sample, all falling within 2 arc minutes of the borders, 
the values at edges were assumed. 


\subsection{The Period--Color Relation}
\label{pc}

The relation between the period and color (PC),
observationally established by Eggen \markcite{egg1} \markcite{egg2}
(1961, 1967),  plays a special role in
studies of W~UMa-type systems. This relation is only moderately tight
as it reflects evolution of some contact systems leading to
longer periods and redder colors for more evolved systems. However, the
short-period/blue envelope of the PC relation is particularly
important as it is expected to be well
defined, being delineated by the least-evolved systems.
Differences in metallicities enter here and might affect the colors
and location of the envelope,
but as was shown in Rucinski \markcite{ruc8} (1995b), the $V-I$
color is relatively insensitive to variations in metallicity,
$\Delta (V-I) \propto +0.04\,[Fe/H]$. For the Galactic Disk stars, the
range in metallicities should be confined to
$-1.0 < [Fe/H] < +0.5$ so that larger
effects might be expected only for strongly metal-deficient systems
(Rucinski \markcite{ruc6} 1994b). Since we have no
independent information on metallicities of individual systems, this
matter will be entirely disregarded here. 

The shape of the short-period/blue envelope (called from now on SPBE)
has been determined on the basis of data for 59 bright, nearby
systems, tabulated by Mochnacki \markcite{moch}
(1985) and transformed to $V-I$ using
the Main-Sequence relations of Bessell \markcite{bes1} \markcite{bes2}
(1979, 1990). The recent improvements of
 the absolute-magnitude calibration (Sec.\ref{cal}) have
led to a confirmation
that the the Main Sequence $B-V$ to $V-I$ transformations are
basically valid for contact binary systems.
It was found that the following simple, two-parameter
approximation describes the SPBE very well:
$V-I=0.053 \times P^{-2.1}$; the period $P$ is in days.  
This expression has no astrophysical
significance and is used here only for convenience. The envelope and
the data are shown in Figure~9.



The period--color relation for the R-sample is shown graphically in
Figure~10, with vertical reddening vectors determined through iteration
described in the next sub-section.
If we concentrate our attention on lower ends of these vectors,
we can see that practically all systems are located substantially
below the SPBE. Obviously, a large fraction of this
shift is due to the reddening, but part of the shift might be also due
to the evolution. We also note that the systems occupy a
much wider range of periods than currently in the standard-system
sample (which becomes unreliable above about 0.6 day), practically up
to the (conventional) limit of one day. The long-period systems
consist of relatively massive stars and are expected to evolve much more
quickly than their late-type analogues so that the un-evolved envelope
is particularly difficult to define for longer orbital periods.
Obviously, we have no information on evolution of individual systems, 
which is actually unimportant for distance determinations
(except that it actually helps seeing distant
systems), but we can approximately correct for the reddening, using
the data on maximum extinction and making an assumption on the extent
of the dust layer.


\subsection{Determination of Absolute Magnitudes and Distances}
\label{det}

Two assumptions on the effective line-of-sight
thickness of the dust layer, $d_0$, 
were used: 8 and 2 kpc. The former assumed
that interstellar absorption increases linearly up to the distance
of the Bulge at $d_0 = 8$ kpc whereas the latter assumed, following
Arp \markcite{arp} (1965) and P94, 
that beyond 2 kpc there is no further absorption. This second
assumption is related to the small thickness of the dust layer in the
Galaxy.  It is equivalent to postulating that at the distance
$d_0=2$ kpc and the galactic latitude of the BW region of $b \simeq
-4^\circ$, the line of sight leaves the dust layer whose
half-thickness at this point is about 150 pc. Note, that the
Sun is supposed to be located about 
20 -- 30 pc above the plane (Reid \&
Majewski \markcite{rei2}
1993, Humphreys \& Larsen \markcite{hum} 1995) 
so that the constraint on the dust layer thickness is
even more stringent.
 
The iteration leading to a simultaneous determination of a distance and
extinction/reddening started with an assumption of $d = d_0$ for each
star and consisted of up to 4 cycles of the following steps 
(always leading to stabilization of $M_I$ to 0.01):
\begin{eqnarray*}
 E_{V-I} & = & E_{V-I}^{max} \times d/d_0 \\
     A_I & = & A_I^{max} \times d/d_0 \\
     M_I & = & M_I(\log P,\, V-I-E_{V-I}) \\
       d & = & 10^{I-M_I+5-A_I}
\end{eqnarray*}

The de-reddened period-color diagrams for both values of 
$d_0$ are shown in Figure~10, where they are identified as R$_8$ and
R$_2$. On the average, the results strongly favor 
$d_0 = 2$ kpc (or perhaps slightly more) as for $d_0 = 8$ 
kpc all systems fall far below the upper envelope. To
explain this behavior in another way, we would have to assume
that practically all systems are intrinsically different than in the
solar vicinity, i.e.\ that they are strongly evolved with down-
and right-ward shifts, away from the envelope of un-evolved systems.
However, assumption of $d_0 = 2$ kpc carries also a danger of
over-correcting the reddening and extinction is some individual
cases. At this time, we feel that we must resort to treating the two
values of $d_0$ as two extremes for the real spatial distribution of
the interstellar dust. We note however that Ng \& Bertelli 
\markcite{ng1} (1996),
have recently presented arguments (using the OGLE data) that $d_0
\simeq 4 - 5$ kpc.

As we will be shown below,
the results of distance determinations for both assumptions
on $d_0$ do differ for intermediate
distances, but the differences are not very large. Also, it
should be stressed that many systems fall outside the strict
applicability of the absolute-magnitude calibration which is marked in
Figure~10 by a broken-line box. This is perhaps most evident for
periods longer than 0.63 days, where not only the calibration is
uncertain, but where the upper envelope for nearby systems is also
practically undefined. 

Inspection of Figure 10 shows that even for $d_0 = 2$ kpc only few
systems fall above and to the left of the SPBE. This can be explained
by absence of very metal-poor systems (with $[Fe/H] < -2$) which are
expected to have their own SPBE, shifted to bluer colors (Rucinski
\markcite{ruc6} \markcite{ruc8} 1994b, 1995b). 

\subsection{Absolute Magnitudes}
\label{abs}

The CMD for the contact binaries for the R-sample and $d_0=2$ kpc, 
in terms of absolute magnitudes $M_I$ and de-reddened colors $(V-I)_0$,
is shown in Figure~11. 
This figure contains also the simplified Main Sequence fits
for Pleiades, as given by Paczy\'nski et al.\markcite{pa1}
 (1994): $M_I = 1.0 + 4
(V-I)$, and for nearby stars, as given by Reid \& Majewski 
\markcite{rei2} (1993): 
$M_I = 1.10 + 4.33 (V-I)$ for $V-I < 0.92$
 and $M_I = 2.89 + 2.37 (V-I)$ for $V-I > 0.92$. 
The contact binaries form a band above the Main
Sequence, which is about 1.5 magnitude wide. The band is somewhat
wider for systems which fall  outside the
applicability range of the calibration (open circles).


We encounter here, for the first time, an
area of concern: A large fraction of the contact systems in
Baade's Window are located in Figure 11  above
the calibrating systems from the solar neighborhood, the latter
tracing well the Main Sequence dependences. 
The location above the Main Sequence is expected for contact binaries
on two accounts: (1)~they are binaries hence they
might show a spread within 0 -- 0.75 mag, (2)~at least a fraction 
among them should show indications of advanced evolution, as
the best explanation for their origin is through
the relatively slow process of magnetic-wind braking with typical time
scales of several Gyr. The system V5 in NGC~188 (V371
Cep), discovered by Kaluzny \& Shara 
\markcite{kal4} (1987) (see also Kaluzny \markcite{kal1} 1990),
might be taken as a prototype for such evolved systems. They should
be, on the average, brighter than other systems and thus should be
visible deeper in space leading to their over-representation in the
OGLE sample. We note that the
numerous faint systems with $I>16.5$ tend to be rather uniformly
spread within the whole length of the colors. An 
obvious over-population of the upper part of the diagram is definitely
due to numerous distant systems, which are in fact very rare in space.

The systematic deviations of absolute magnitudes could be also due to
application of 
wrong reddening corrections, as the individual values of
$M_I$ depend on the reddening-corrected color. However, the sense of
the discrepancy is wrong. 
Figure~11 shows the results for the preferred by us,
the shorter dust length 
scale of $d_0=2$ kpc. For the longer scale of $d_0 = 8$ kpc, the
discrepancy becomes larger and a gap opens up between the Baade's
Window data and the standard-system data. 
 Since for the larger $d_0$ the reddening corrections are on the
average smaller (the reddening climbs slower with distance), 
these results would imply that even for $d_0=2$ kpc, the
reddening corrections are systematically slightly too small. This is
inconsistent with the assumption that that data in S96 
gave maximum values of the reddening. 

Unfortunately, we have no way of
disentangling color shifts due to evolution from those due to
reddening. In particular, we cannot attribute the whole deviations in
$M_I$ to incorrect reddening, as solutions of such an
 inverse problem are
unstable and lead to a very large spread in $E_{V-I}$. In this
respect, the maps in S96 give us at least a way to confine 
the values of reddening within reasonable limits. 
The other possibility is that the $V-I$ colors contain a small
systematic error (in the sense of being too red) for faint stars. 
We discuss such a possibility in Section \ref{problem}
Here we note that a simple allowance for such an error
of about 0.2 mag in $V-I$ does not change our conclusions, in that the
deviations in $M_I$ between Baade's Window and calibrating systems
become only slightly smaller.

The data in the CMD are subject to truncation from below and from
above in absolute magnitudes. The upper boundary at about $M_I \simeq
1$ (but dependent on color) is related to the
conventional limit on the orbital period of one day. It is shown by a
dotted line in the figure which was obtained by inserting $P = 1$
day into the absolute-magnitude calibration. The lower boundary
is obviously related to the magnitude limit of the sample at $I \simeq
17.9$.  From the discussion of distances (the next sub-section) we have
indications that the sample is probably complete to about 3 kpc. With the
average extinction in the field, the corresponding absolute-magnitude
limit would be then at $M_I \simeq 4.6$.
 

The luminosity functions in $M_I$,
based on systems to 2 and 3 kpc are shown in
Figure~12. The faint end is by one magnitude deeper for the smaller
volume, but the statistics is obviously more limited.
The same functions, transformed from
the $M_I$ to $M_V$ magnitudes are shown also in the lower panels of
the same figure. The $M_V$ functions are
blurred by a spread in $V-I$, but have an advantage that they
can be compared directly with the LF's for other objects. In
particular, when we compare them with the luminosity function
 for Blue Stragglers in globular clusters (Fusi Peci et al. 1993,
Sarajedini \markcite{sar} 1993), 
which peaks at $M_V \simeq 3$ and extends between $2 < M_V
< 4$ with strongly tapered ends, we see that, on the average,
 the contact binaries are fainter than the globular-cluster
Blue Stragglers so that only the relatively 
weak bright tail of the LF might contain a BS contribution.


\subsection{Distance Distribution}
\label{distr}

Three directly 
observed quantities, $P$, $I_{max}$ and $V-I$, two values found
by interpolations in the 
maps, $A_I$ and $E_{V-I}$, and one derived value,
the distance $d$, can be related by combining our calibration with the
standard expression linking observed and absolute magnitudes:
$$I=b_{P(VI)}\,\log\,P+
b_{VI}\,(V-I)-b_{VI}\,E_{V-I}+b_{0(VI)}+5\,\log d-5+A_I$$
In simplified relations below we normally set $A_I =
1.5\,E_{V-I}$, with $E_{V-I}=0.6$, but for individual systems the
iterative process described above was used. We should note that
accuracies of distance determinations depend practically only on
uncertainties in magnitudes, colors and values of
reddening as, for the present purpose, the
 periods are known practically without errors. Thus, the respective
contributions to the distance errors $\epsilon \log d$, 
from errors in magnitudes, colors and reddening  would be:
$\propto \epsilon I/5$,
$\propto b_{VI}\,\epsilon (V-I)/5 \simeq 0.46\, \epsilon (V-I)$, and
$\propto (1.5-b_{VI})\,\epsilon E_{V-I}/5 \simeq -0.16\, \epsilon
E_{V-I}$. Because of the size of 
the color coefficient $b_{VI}$ in the absolute-magnitude calibration,
the dependence on any possible  systematic color deficiency
in the calibration of the OGLE photometric system is about two times
smaller than it would have been for the standard technique
of absolute magnitudes estimated from the Main-Sequence
fitting.  In particular, for a systematic error of
$\epsilon (V-I) \simeq 0.1$, the distances would be systematically
wrong by $\epsilon \log d \simeq 5$\%.


Results of the determinations of the distances for the R-sample and
$d_0=2$ kpc are 
shown in Figure~13. The upper panel shows the data in the 
period--distance plane. In principle, the two quantities should
not correlate as -- barring really
unusual and then extremely-interesting astrophysical causes -- 
the detection rate should be the same at all
distances for a given period. 
In fact, we clearly see lack of short period, low
luminosity systems beyond about 2.5 -- 3 kpc, as expected given the
magnitude-limited nature of the sample. This way, numerous
short-period systems which dominate the sky-field sample, with a peak
at about 0.35 day and a sharp cutoff at 0.22 day (Rucinski 
\markcite{ruc3} 1992), are
entirely eliminated for distances beyond about 3 kpc.
At larger distances, rarer, long-period systems dominate in numbers.
They do not show any tendency for a decrease in numbers beyond about
2.5 -- 3 kpc, and seem to be visible all the way to the Bulge, with a
few systems -- admittedly outside the calibration applicability by
about 1.5 mag -- located in the very Bulge at 8 kpc. 
It is interesting to note a hint of a small
gap between 6.5 and 8 kpc. Up to about 6.5 kpc, the
distribution in logarithmic units of the distance
climbs approximately linearly without showing
any structure, indicating approximately flat distribution in the 
distance. 

All 9 systems which seem to be located at the distance of the
Bulge have well defined light curves
and are prime candidates for a difficult, but extremely important,
radial-velocity follow-up study\footnote{Two systems
formally beyond 8 kpc are both questionable:
\#0.198, has a poor light curve, whereas \#6.121 has an unusually blue
color (cf.Section \ref{cmd}) so that its distance might be entirely
erroneous.}. The systems in the Bulge have designations: \#1.121,
\#1.138, \#1.199, \#2.132, \#3.094, \#3.142, \#3.171, \#3.178,
\#7.163.
Except \#1.121 and \#3.094 which have $16.5 < I < 17$, all
are fainter than $I = 17$. This location is very close
to the Turn-Off Point of the Bulge, which is now placed
by Kiraga et el. \markcite{kir} (1996) at about $I = 18$. 
In the CMD in Figure~7, all nine 
appear in the ``corner'' just below the Disk MS, with $V-I
\simeq 1.1 - 1.3$. Some of these nine systems show unequal minima.

The curved,
short-period cutoff in the distribution of distant systems can be
explained entirely by the existence of the SPBE
on the period--color relation. In order to be visible from large
distances, a system must be blue and must have a long
period. Both quantities are constrained by the envelope. The upper
panel of Figure~13 shows the SPBE, as in Figure~10, transformed to
the distance -- period plane using the average extinction $A_I = 0.9$,
for two values of $I$, 17.9 and 17.0. The un-evolved
envelope of the period--color relation does indeed
set a strong limit on observability of distant contact binaries.


Our determinations of distances have been based on a more likely
assumption on the spatial
distribution of reddening, with an abrupt stop in its increase at
$d_0=2$ kpc. Has this assumption influenced the distance
determinations to a large degree? The results for the opposite extreme
of $d_0=8$ kpc, designated as R$_8$, are shown in Figure~14.
The distance distribution is generally flatter
than for R$_2$, with largest changes taking place for intermediate
distances. A weak depression in increase of numbers with distance at
about 3 kpc seems to be somewhat stronger than before, but -- in
general -- the distributions of distances are similar.

The results presented in Figures~13 and 14 seemed to the author
 so important that he could not resist a temptation 
to turn back to the whole O-sample and repeat the whole procedure
again (remembering that some systems of this sample might be less
carefully screened for quality of data and classification than those
of the R-sample). The results are shown in Figure~15, for $d_0=2$
kpc. They do not
differ much from those for the R-sample. The small subset of
systems in the Bulge seems to be even better defined, forming a
vertically clustered grouping at $d \simeq 8$ kpc in the upper panel.


The relation between the distances
and intrinsic colors for the R-sample is presented
in Figure~16. We see a large range of intrinsic colors
reaching quite red values for a few systems, but
most are confined to a moderately narrow range between 
$0.4 < (V-I)_0 < 1.0$. The three systems which
are intrinsically very red with $V-I>1.5$ are: 
\#3.053, \#7.112 and \#8.072. At large distances, there
exists an obvious limit at red colors related
to faintness of red, short-period systems. On the blue side of the
diagram, 
we do not see any influence of the selection effects related
to the SPBE, since colors as blue as
approaching $(V-I)_0 \simeq 0.2$ should be permitted by the
envelope.  Thus, there exist no contact systems in the OGLE sample bluer
than $(V-I)_0 \simeq 0.4$.


Having the determinations of
distances, we can return to the color--magnitude
diagram for the whole sample and consider positions of systems in the
3-D space of observed $V-I$, $I$ and distance. This is shown in
Figure~17 where distances are coded by sizes of symbols. 
As expected, distant systems cluster in the lower part of
the diagram, with a striking excess in the ``corner'' below $I > 16.5$
and within $1.1 < V-I < 1.4$ or $0.5 < (V-I)_0 < 0.8$. The selection
effects operating here are related to two limitations of the sample. 
One of these is due to the cut-off of the sample at one day.
For any distance, slanted straight lines (parallel to the two
straight ones for 4 and 8 kpc), define two different
volumes: Above the line, we see a mixture of
systems at different distances, but below this line, systems
 must have periods {\em shorter\/} than 1 day to be detectable 
between us and the assumed distance. There exists also a limit on the
period on the short side, which results from the period--color relation, as
not all combinations of periods
and colors are permitted, both quantities being constrained by the
SPBE of un-evolved systems in the
period--color relation (cf. Section \ref{pc}).
These two limitations produce -- for each distance --
 a region on the CMD diagram where contact
binaries might be observed. Two such regions are shown 
in Figure~17 for the
distances of 4 kpc and 8 kpc. Similar ``catchment
areas'' can be defined for other distances, by simple vertical shifts 
on the CMD. The areas do not impose
limits on the presence of blue systems in the sample so that such
systems do not apparently exist at large distances.


Finally, the distribution of the 
systems in the period -- magnitude diagram
(Figure~18) shows a tendency for clustering  at faint
magnitudes and long periods. This tendency can be
identified with the increase in numbers of systems at large distances,
close to the limits of our sample,
as the observed clustering tendency approximately follows 
lines of equal distance and approximately equal
color. It is gratifying to see this link to the distances as it
indirectly proves that our interpretations are basically correct. 



\subsection{Spatial Density and Apparent Frequency of W~UMa-type systems}
\label{freq}

As was shown in the previous section, the sample of contact systems
within the ``pencil-beam'' view of the OGLE search is probably
complete to 3 kpc. To see the sensitivity to this assumption, we will
consider two distances,  2 kpc
and 3 kpc, as limits for completness. For
$I_{max} = 17.9$ and assuming for simplicity (but inaccurately)
 the same absorption of $A_I = 0.9$, the limiting absolute magnitudes
would be $M_I = +5.5$ and $M_I = +4.6$, respectively.
Within the volumes defined by the two limiting distances, 
there are 27 and 98 contact systems in the
R$_2$-sample and 39 and 141 systems in the O$_2$-sample. The {\it
apparent\/} density of contact systems can be simply evaluated by
dividing these numbers by the volume of
the spatial cone defined by the OGLE field. For the two limiting distances,
one obtains then $(7.0 \pm 1.4) \times 10^{-5}$ and $(7.6 \pm 0.8)
\times 10^{-5}$ stars/pc$^3$
for the R-sample, and $(10.2 \pm 1.6) \times 10^{-5}$ and $(10.9 \pm
0.9) \times 10^{-5}$ stars/pc$^3$ for the O-sample, where the errors
come from the Poisson statistics.
We note that the estimates for 2 kpc and 3 kpc are very
similar indicating completness of the sample. This should not be
suprising as the contact binaries show a short-period, red-color
cut-off so that faint systems practically do not exist. While the
densities are well defined for both distances, the essential
uncertainty here comes from the initial definitions of the R- and
O-samples. 

The density of the contact systems is apparently high. In
particular, if we multiply the above estimates 
of the apparent density by 2 times, to correct
for undetected systems of low orbital inclinations, and compare the
resulting {\it spatial\/} density with
the local density of RR~Lyrae stars, we see that the contact binaries
outnumber the latter by a huge factor. The local density of all RR~Lyr
stars is $(6.2 \pm 1.4) \times 10^{-9}$ stars/pc$^3$ (Suntzeff, Kinman
\& Kraft \markcite{sun} 1991), so that for a round number of
$15 \times 10^{-5}$ contact systems per cubic parsec, we obtain
the ratio of 24,000 times. Obviously, this number is applicable only
to the solar neighborhood as the RR~Lyrae stars show a strong
galacto-centric concentration.

While the density of the W~UMa-type systems in the current sample can
be relatively easily evaluated, the relative frequency requires
assumptions on the sample of normal stars to use as reference. 
It should be remembered that contact systems with spectral types later
than about K2--K5 do not exist, a limit apparently related to the
existence of the sharp cutoff in the period distribution. This range
of the spectral types coincides with the region where the luminosity
function for normal stars climbs steeply, so that the 
results on the relative frequency
must depend on the limiting distance or limiting absolute
magnitude assumed for normal stars. By integrating the stellar density
function of Wielen, Jahreiss \& Kr\"uger \markcite{wie}
(1983) to limiting magnitudes
of our sample (which gives about 
0.01 stars/pc$^3$) we obtain a very high spatial
frequency of of W~UMa systems of about 0.015 or one such a system per
about 67 Main Sequence stars. However, this number critically depends on
how far is the luminosity function integrated in its rising part, and
carries uncertainty of about two times. It
is actually easier to relate the number of W~UMa systems to the {\it
total\/} number of stars by integrating the stellar density function
all the way to very small stars. Here we used the extension of the luminosity
function to faint M-dwarfs by Gould, Bahcall \& Flynn 
\markcite{gou} (1996) which
drops sharply for faint stars. Such integrated total number density
of 0.06 stars/pc$^3$ is well defined. The absolute spatial
frequency of the W~UMa systems is then 
0.0025, or one such a system per 400 stars, including M-dwarfs.

The relative frequency of the contact systems can be also evaluated
using the OGLE data for normal stars.
Distributions for single stars in Figures 7, 8 and 16
were based on what we called in Section \ref{cmd} the ``CMD-stars'',
i.e.\ stars with high-quality photometry which could best define the
color--magnitude diagrams. The databases for these stars were easily
accessible from the {\em ftp\/} sources, as described in P94, so that
these stars were the main basis for relating the contact binaries to
normal stars. However, these
CMD-stars are not useful for statistical analyses,
because their numbers were biased in progressively fainter
magnitude bins in a different way than stars which were analyzed for
variability.  For statistical applications, the full count of
all stars observed and analyzed in the OGLE project would be 
necessary. Such a  full catalog 
is not available at this moment. However, the first
installment of the full catalog for the Baade's Window
central field has just been
published (Szyma\'nski et al. 1996). As described in this paper,
statistics of faint stars are strongly biased in the CMD sample, 
with the ratio of all stars to the CMD stars reaching a large
factor of 1.90  for the cululative count of stars with $I < 18$. 
Using the new central-field data available over the
{\em ftp\/}, we determined this ratio in magnitude intervals
for the range of $14 < I < 18$. In steps of one magnitude, it changes
approximately as: 1.05, 1.35, 1.8, 2.15, 1.6, and peaks at $I \simeq
16.5$ reaching 2.2. The smoothed version of
this ratio was applied as a correction factor for all 9 fields,
in an estimate of the total number of stars that 
had undergone scrutiny for variability by the OGLE Project.

To estimate the total number of Main Sequence stars
to the distances of $d=2$ and 3 kpc, 
we counted all CMD-stars in all nine fields 
with $V-I < 1.5$ (to avoid contamination by
the Red Clump giants), which fell 
above the Main-Sequence lines given by P94,
$I = 1.0 + 4\,(V-I)_0 + A_I + 5 \log (d/10\,{\rm pc})$, for the assumed 
values $A_I = 0.9$, $E_{V-I}=0.6$. These numbers were corrected by the
magnitude-dependent factors (as described above), 
giving the expected numbers of the Main Sequence stars
with $V-I < 1.5$ equal to 10065 and 33055, to 2 and 3 kpc,
respectively. Using the numbers of the contact systems, as given at
the beginning of this section, we obtain
the apparent reciprocal frequencies (in the sense of numbers of ordinary
dwarfs per one W~UMa-type binary), and for both distances:
$373 \pm 72$ and $337 \pm 34$  for the R-sample and
$258 \pm 41$ and $234 \pm 20$  for the O-sample. 
Remembering that the O-sample might contain
mis-classified systems of other variability types 
and the R-sample might be overly restrictive,
we can say that in round numbers the apparent reciprocal frequency is
250 -- 300. This frequency compares
very well with similar estimates for the old open clusters and their
background/foreground fields 
 of $275 \pm 75$ and $285 \pm 120$ given in R94.

Our current estimates of the apparent frequency of the W~UMa systems
are substantially higher than estimated by Duerbeck \markcite{dur} (1984)
for nearby systems at about 1000, confirming our earlier supposition
(Ka\l u\.zny \& Rucinski \markcite{kal3} 1993,  
Rucinski \& Ka\l u\.zny 1994 \markcite{ruc10})  that the sample
of bright, field systems is incomplete, being
biased against low-amplitude systems.
 The implied {\it spatial\/} reciprocal frequency based on the OGLE
data is approximately 125 -- 150,
assuming that undetected
systems with low inclinations contribute one half of all
systems. This result is consistent with the direct estimate based on
stellar densities, at the beginning of this section, of about 67 Main
Sequence stars per one contact system which can be evaluated only to a
factor of about two times.


\subsection{The local sample to 3 kpc}
\label{local}

The luminosity functions
based on the volume-limited samples to 2 and 3 kpc have been
discussed in Section \ref{abs}. Here, we
look at the statistics of the two most important,
time-independent quantities, the orbital periods and intrinsic colors.
The histograms of both quantities for the R$_2$ sample, limited to $d <
3$ kpc, are shown in Figure 19. 


The upper panel of Figure 19 shows the distribution of orbital
periods. It is very similar to such a distribution for field contact
binaries (Rucinski \markcite{ruc3} 1992),
which shows a peak at short periods, with a sharp cut-off close to 0.25
day and a gentle slope for longer periods. Apparently, fears expressed
by this author that the
field-system distribution might be affected by observational
selection effects were premature. Potentially, this histogram should 
contain very important information on the formation and structure of
the contact system. We note that most of its features do not have good
and un-equivocal explanations. 

The intrinsic-color distribution in the lower panel of Figure 19
shows a broad maximum between $0.45 <
(V-I)_0 < 1.05$ with the total extension between $0.2 < (V-I)_0 <
1.5$. There exist also a few systems with very red colors, reaching
$(V-I)_0 \simeq 2.5$. As we comment in Section \ref{pop}, the
concentration of colors is reminiscent to that observed for Turn-Off
Point stars of old population, although the distribution for contact
binaries is wider and extends further to the red.

\subsection{Possibility of an error in the color calibration}
\label{problem}

In Section \ref{abs} concerns have been expressed about 
 a small but systematic difference in average
absolute magnitudes between the Baade's Window systems and nearby
systems used for our calibrations. This difference might indicate a
problem with the color scale in the OGLE data
for faint stars. In fact, Kiraga, Paczy\'nski and
Stanek \markcite{kir} (1996) have recently pointed out 
some evidence for lack of consistency between the color-magnitude
diagrams in Baade's Window as available from OGLE (Udalski et
al. \markcite{uda1} 1993, Paczy\'nski et al. \markcite{pa1} 1994a), 
from the HST (Holtzman et al. 
\markcite{hol} 1993), and the color --
magnitude diagrams for the globular clusters 47 Tucanae 
(Ka\l u\.zny \markcite{kal2} 1996)
and the open cluster NGC 6791 (Ka\l u\.zny \& Udalski  
\markcite{kal6} 1992).  A possible
reason for the inconsistency is a systematic shift in the OGLE $V-I$
colors for the stars fainter than the ``Red Clump''. 
The inconsistency would manifest itself for  $I > 15$
where the listed $V-I$ colors in the OGLE data seem to become 
too red, with the discrepancy reaching about 0.1 in $V-I$
for $I = 17$, and stabilizing at this value for fainter stars. 
This is just a possibility at this time, and it may
be caused by a non-linearity of the CCD detector (cf. Udalski et
al. \markcite{uda1} 1993) or, partly, by blending, 
with the red stars below the detection threshold affecting in a small,
but systematic way the stars above the detection threshold.
This issue shall be clarified in the near future using 
independent photometry based on a different CCD detector (Ka\l u\.zny
\& Thompson, \markcite{kal5} in preparation).
 
To estimate the influence of such an error, the data have been analyzed
with colors artificially changed following the above suspected 
dependence of the error on brightness, 
by assuming that the discrepancy increases linearly within $15 < I <
17$ to 0.2 and then stays constant. Even for such an exaggerated case,
the changes are small. The period -- color
diagram in Figure 10 is almost un-affected, although the gap in the
distribution of points and the SPBE less evident, especially for
the R$_8$ case. This is as expected, as faint systems, which would be
mostly affected by the color error, have --
on the average -- longer periods (cf. Figure~18). Further
experiments show that the discrepancy
between the absolute magnitudes of the OGLE systems and nearby systems
in Figure 11 remains for artificial data, although the
deviations do become smaller. The relatively
small change can be explained by our
scheme of the coupled absolute-magnitude, distance, reddening
determination which distributes color differences in a more complex
way than just a simple shift $\Delta M_I \propto b_{VI}\,\Delta (V-I) 
\simeq 2.3\,\Delta (V-I)$.
The largest differences appear in the
distance -- period relations (Figures 13 -- 15): The data points tend
to be more spread at large distances, there is no concentration of
points at 8 kpc, and several systems are moved to distances as large
as 10 kpc. The gap between 6.5 -- 8 kpc is filled. The 
distance -- color relation in Figure 16 obviously shows the same
features in terms of distances, but the concentration in intrinsic
colors around the TOP colors, which was 
discussed in Section \ref{distr} remains. 

Thus, even if an error in the color is present, we can uphold
our main conclusion of the paper on the global distribution of contact
binaries in space, but some details would have to be
modified. Also,  since the error may appear only at faint
magnitudes, the results on the local frequency of contact binaries
should remain entirely un-affected. 

\section{POPULATION CHARACTERISTICS OF THE CONTACT BINARIES}
\label{pop}

Two mechanisms of formation of contact binaries
from detached binaries have been proposed. One involves
the magnetic-wind evolution of the orbital angular momentum 
(Van't Veer \markcite{vv} 1979, Vilhu \markcite{vil}
1982, Guinan \& Bradstreet \markcite{gui} 1988)
and seems to be the main route to produce such systems 
in open clusters and in the field. The second mechanisms utilizes
dynamical interactions in dense cluster cores, possibly involving 
binary systems (Leonard \& Fahlman \markcite{leo1}
1991, Leonard \& Linnell \markcite{leo2} 1992),
and is expected to contribute mostly in dense stellar clusters.
Of importance for distinguishing between these two mechanisms would be
data on population characteristics of the contact binaries.  
Although we have scant data on specific ages of contact binaries 
and we have no idea about the distribution of the ages, 
there are indications that they
are quite advanced. This is suggested by space motions of the nearby
systems (Guinan \& Bradstreet \markcite{gui} 
1988), which correlate weakly
with $\delta (U-B)$
metallicity indices (Rucinski, unpublished), as well as by the recent
detections in old open clusters. 
There exist single contact binaries in Praesepe and
Be~33 with ages of 0.7 -- 0.9 Gyr, but then
their numbers increase for older
clusters with several systems in the old clusters NGC~6791, NGC~188,
Be~39 and Cr~261 (Ka\l u\.zny \& Rucinski \markcite{kal3} 1993, 
Rucinski \& Ka\l u\.zny \markcite{ruc10} 
1994, Mazur et al. \markcite{maz} 1995). 
Of particular importance is the fact that contact binaries seem to
exist in old open and globular
clusters on both sides of the Turn-Off Point (TOP). 
First W~UMa systems in globular
clusters were found with surprisingly high frequency
among Blue Stragglers (Mateo \markcite{mat1} 1993, 
\markcite{mat2} 1996), 
but recently Yan \& Mateo \markcite{yan} (1994) 
discovered them also below the TOP (but above and along the MS) in
M71. Observational selection effects are formidable below the TOP
in globular clusters in most cases so that it would be premature to
conclude that the MS contact systems do not exist there. 
However, the first data
from the Hubble Telescope (Rubenstein \& Bailyn \markcite{rub}
1996, Edmonds et
al. \markcite{edm} 1996) indicate that globular clusters might, indeed,
 be deficient in Main Sequence contact systems.

An open cluster with systems on both sides of the TOP is 
the populous old open cluster Cr~261 (Mazur et al. \markcite{maz} 1995):
For some 12 certain and 6 uncertain members of the cluster (among 28
contact systems in the field), 4 are Blue Stragglers. Not only this is
the only cluster unambiguously showing W~UMa systems on both sides of
the TOP, but its contact 
Blue Stragglers seem to have another property: their
amplitudes are systematically smaller than for the rest of the
systems. If this is not a statistical coincidence due to the small sample, 
it excellently links with the suggestion of Eggen \& Iben 
\markcite {egg3} (1989) that
the extremely small mass-ratio system, AW~UMa, is a Blue Straggler in
the field (for references to other possible cases, cf. Mateo 
\markcite{mat1} 1993). 

Although we have no radial velocity or abundance information for
the contact binaries in Baade's Window,
their spatial distribution and relative frequencies, as well as
preference of colors to group within a range around $0.4 < (V-I)_0 <
1.0$ (Section \ref {local}) shed much light
on their population characteristics. Apparently, we see contact
binaries at all distances without any indication of any structure in this
distribution; they do not seem to be concentrated to the
Galactic plane. These characteristics are consistent with an old
Galactic population. The concentration of colors to a relatively
narrow interval is reminiscent of the enhancement observed as a broad
vertical band on color--magnitude diagrams of the
stellar field (Gilmore \markcite{gil}
1990, Reid \& Majewski \markcite{rei2} 1993), in the range of
``vertical'' evolution in the TOP region. In the old
population, there are practically no bluer stars than about
$B-V \simeq 0.4$ or, equivalently,
$V-I \simeq 0.45 - 0.5$, but then, within $\Delta (B-V)
\simeq 0.2$ or $\Delta (V-I) \simeq 0.25$,
 the density is substantially enhanced by stars of the TOP
region. Obviously, the actual location of the enhancement depends on
the reddening in the particular direction. Expressed in the
de-reddened colors, the enhancement for the contact
binaries seems to be wider than the normal width of
$0.4 < (V-I)_0 < 0.7$; in Figure~19 it extends to the red to  
$(V-I)_0 \simeq 1.0 - 1.2$ or even further. While this extension can be 
perhaps explained through changes in stellar structure during the
interaction of components when establishing the contact, one
of the main goals for the future would be to establish whether the red
cutoff is real and does not result from a selection effect or another
bias in the photometric material. In
Figures~16 and 19,
some systems have surprisingly red colors, so that we
directly see that contact
binaries are not confined exclusively to the TOP region.

Having established that contact binaries belong to an old Galactic
population, we must pose the question: 
Which old population is it? It seems unlikely that
contact binaries in their majority are as old as
stars of the Halo or the
Thick Disk. According to Gilmore \markcite{gil} 1990 and Reid \&
Majewski \markcite{rei2}
1993, the Extended/Thick Disk population has scale heights of
about 1400 -- 1600 pc. The line of sight to Baade's Window passes
the Galactic Center at the distance of about 560 pc, well within these
scale heights, so that uniform density of the W~UMa systems
in space would agree with characteristics of this population.
 The main problem is with the low local density of such 
very old stars of only about 2\%. The apparent frequency of the contact
systems found here suggests a spatial density of about 1/150 or more, 
or 0.7\%, which would require that about 1/3 of these very old
stars were contact binaries. This is difficult to accept.
A much more likely parent population of the contact systems is the Old
Disk of solar-age stars, 
with the scale height of about 325 pc. These stars dominate in
numbers in the solar vicinity and are the most likely donor of close
binaries for formation of contact systems. A hint of a break in the
continuity of the distance distribution at about 6.5 kpc might be
actually due to the line of sight leaving that disk at $z \simeq 420$
pc.

The current data cannot unfortunately shed light on the contribution of
Blue Stragglers to the contact binary sample. Possibly, a careful
search in our database
for low-amplitude, totally eclipsing systems similar to AW~UMa
could lead to isolating a sample likely to consist of Blue Stragglers.
We should note that if the
nine systems which we see in the Bulge do belong to the oldest Galactic
population, some of them might be actually Blue Stragglers as they are
all relatively blue. However, even the
bluest among them would be then relatively ``mild'' BS's, 
as their colors are only moderately different from those expected for
the Bulge TOP: the  observed colors, $1.12 < V-I <
1.44$, and the intrinsic colors, $ 0.49 < (V-I)_0 < 0.78$.

\section{CONCLUSIONS}
\label{sum}

The paper presents discussion of properties of contact systems
discovered in 9 fields monitored during the micro-lensing project
OGLE. A simple automatic classifier based on the Fourier analysis of
light curves permitted isolation of 388 well observed
contact systems with available $V-I$ colors from among 933 eclipsing
systems in the OGLE Periodic Variable Star Catalog. Utilization of the
$M_I=M_I(\log P, V-I)$ calibration permitted determination of
distances in an iterative scheme involving maximum values of reddening
and extinction taken from the maps of Stanek \markcite{stan1}
(1996). Analysis of the
distance distribution reveals no discontinuities or unexpected
features, except for those which can be 
at least partly explained by an interplay of known selection effects. 
An apparent lack of contact systems 
bluer than $V-I \simeq 1.1 - 1.2$ at large distances seems to be a
genuine feature of the color -- magnitude diagram. Most of the systems
appear in the color range $0.4 < (V-I)_0 < 1.0$, i.e.\ the blue edge of
the distribution is the same as for stars of the old population which
concentrate in the Turn-Off Point (TOP) 
region, but the red edge appears at redder colors.

Contact binaries with periods longer than one day, which
are intrinsically brighter than the W~UMa-type variables, but which
normally are not considered together with the W~UMa-type systems
will be the subject of a separate investigation. No
absolute calibration is at present available for those long-period
systems. We note, however, that the calibration for the W~UMa-type systems
within $0.6 < P < 1.0$ day is really an extrapolation and should
be properly established. Also, due to its particular importance, the
short-period/blue envelope (SPBE) for the un-evolved systems, 
in the period--color diagram, urgently needs a better
definition. These progresses can come only with an increase in
quality and availability of
photometric data for contact binaries, which are still lagging behind
availability of light curves.

A first attempt to construct luminosity functions for volumes limited
by distances of 2 and 3 kpc, where selection effects should be small
was made. These functions are truncated at $M_I = 5.5$ and $M_I =
4.6$, respectively by the magnitude limit of the sample. This
truncation makes it impossible to firmly establish 
whether there exists excess of contact binaries in the TOP region,
whose presence is indicated by the distribution of intrinsic colors
for the whole sample. Extension of the luminosity functions to
fainter magnitudes should be one of the primary goals for the future.
The numbers of stars contained in the volumes used for the luminosity
functions permitted an estimate of the frequency of 
contact binaries. The apparent frequency equals to one contact system per
about 250 -- 300 Main Sequence
stars. This estimate, with an approximate correction for undetected,
low-inclination systems, leads to the spatial frequency of about 1/125
-- 1/150. 

A direct evaluation of the spatial density of the contact
systems of about $1.5 - 2.0 \times 10^{-4}$ per cubic parsec leads to
an even higher frequency when comparison is made with Main Sequence
stars of similar absolute magnitudes. While the density estimate is
quite robust, the relative frequency estimate obtained that way
is uncertain as it
depends strongly on the assumed magnitude limit in the rising part of
the luminosity function for normal stars. 

The high spatial frequency of the W~UMa-type systems, together with the 
 uniform spatial distribution all the way to the Bulge and with the
 preference for intrinsic colors to cluster within 
the $(V-I)_0$ interval of 0.4
to 1.0 suggest that the contact binaries 
belong to the old stellar galactic
population. Considering predominance of the Old Disk
of solar-age stars in the solar neighborhood over the
Extended/Thick Disk or Halo populations, it is most likely that the
Old Disk stars are the main donor of contact binaries.
The relative contribution of the contact
Blue Stragglers to the sample cannot be presently evaluated. However,
analyses of individual light curves might help in isolating a
sub-sample of very low mass-ratio systems, possibly dominating among 
the contact Blue Stragglers.

\acknowledgments
 
The author would like to express thanks to the OGLE Project for access
to their database and to Hilmar Duerbeck and Yuen Ng for useful comments. 
 Special thanks are due to Bohdan Paczy\'nski,
Andrzej Udalski and Krzysztof Stanek for their
expert help with various technical and interpretative
details of this work. 

This work would be impossible without help of
Bohdan Paczy\'nski and of my wife Anna.

The research grant from the Natural Sciences and Engineering Council
of Canada is acknowledged with gratitude.


\begin{figure}           
\centerline{\psfig{figure=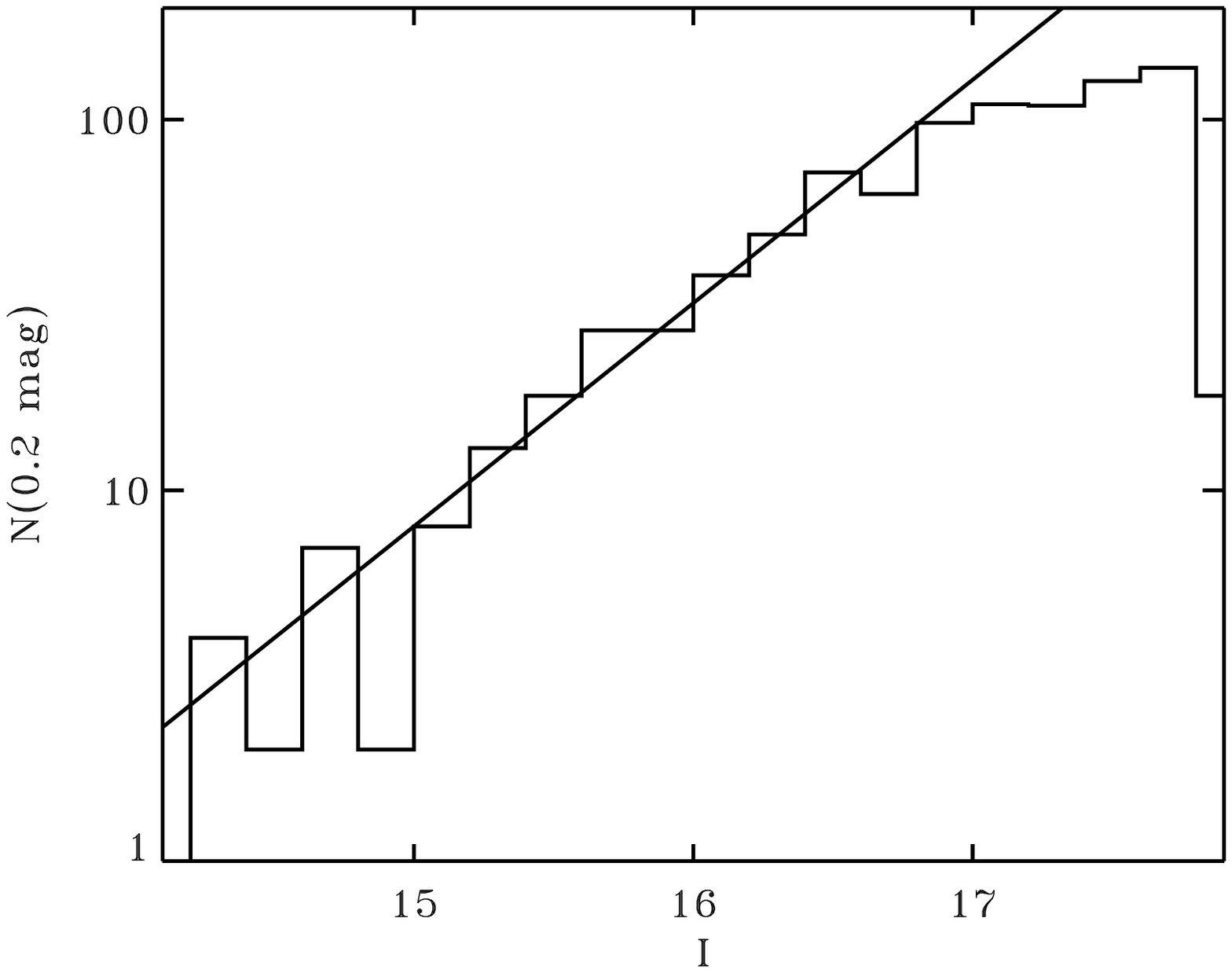,height=2.5in}}
\vskip 0.5in
\caption{Histogram of $I_{max}$ magnitudes for
all 933 eclipsing systems in the Baade's Window OGLE sample. The line
gives the $N \propto 4^{I-14}$ dependence.}
\end{figure}

\begin{figure}           
\centerline{\psfig{figure=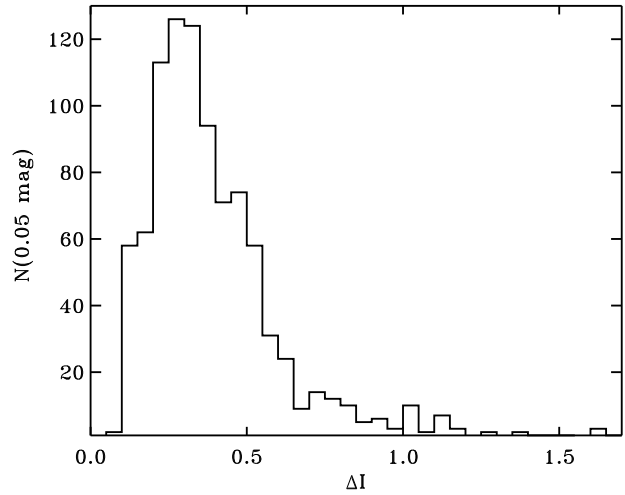,height=2.5in}}
\vskip 0.5in
\caption{Histogram of variability amplitudes $\Delta I$ for
all 933 eclipsing systems of the sample.}
\end{figure}

\begin{figure}           
\centerline{\psfig{figure=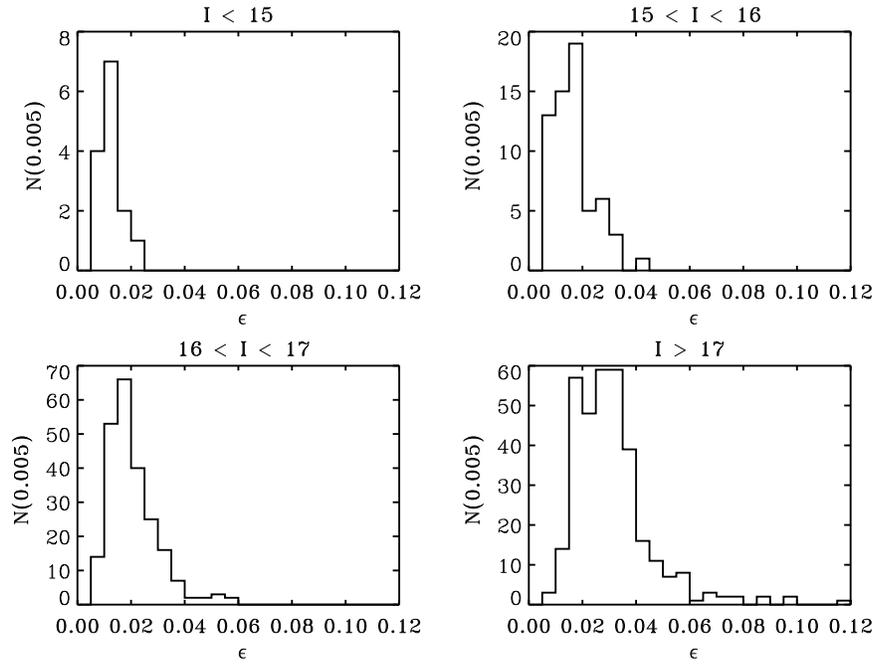,height=3.5in,angle=90}}
\vskip 0.5in
\caption{Histograms of observational errors for all eclipsing systems of
the sample, in four brightness bins, as indicated in the figure. The
errors are in light units, which for small $\epsilon$, are
 close to those in magnitudes.}
\end{figure}

\begin{figure}           
\centerline{\psfig{figure=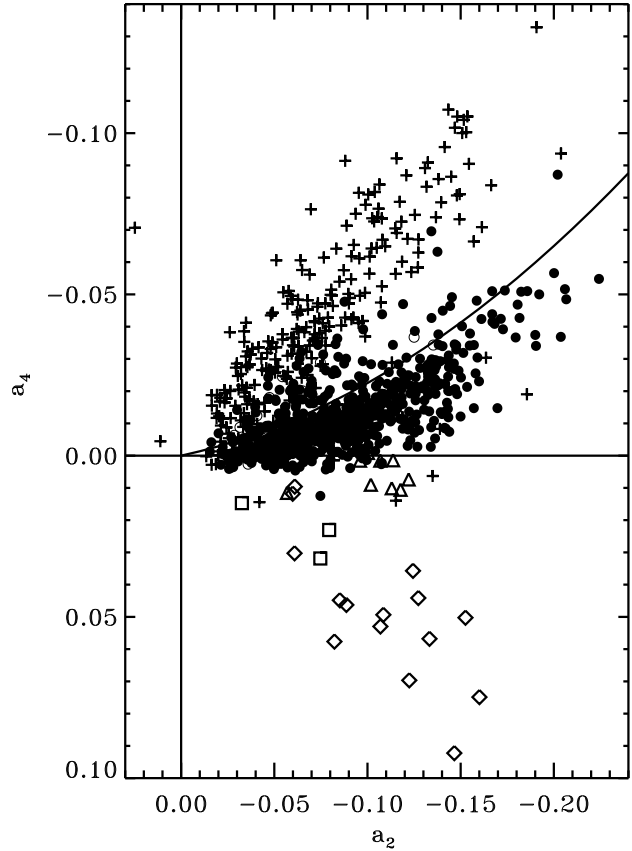,height=4.5in}}
\vskip 0.5in
\caption{The even Fourier coefficients, $a_2$ and $a_4$, are plotted
here for all 933
eclipsing systems of the OGLE sample. They fall in the upper part of
the diagram and can be thus distinguished from pulsating stars
mistakenly taken for contact binaries, which appear in the lower part.
As an illustration, pulsating
stars in one OGLE field (BWC) are included here with periods equal
to twice their real periods. The symbols give original classifications
in the OGLE Catalog: EW systems are marked by filled circles; the EB systems
are marked by open circles, and all EA and E systems as well as all
those that had uncertain classifications (question mark with the
class) are marked by crosses. The continuous curve gives the 
envelope for the inner (marginal) contact, as discussed in the text. The
pulsating stars in BWC are identified by open symbols:
rhombuses for RRab-type, triangles for RRc-type and squares for
SX~Phe-type variables. 
}
\end{figure}

\begin{figure}           
\centerline{\psfig{figure=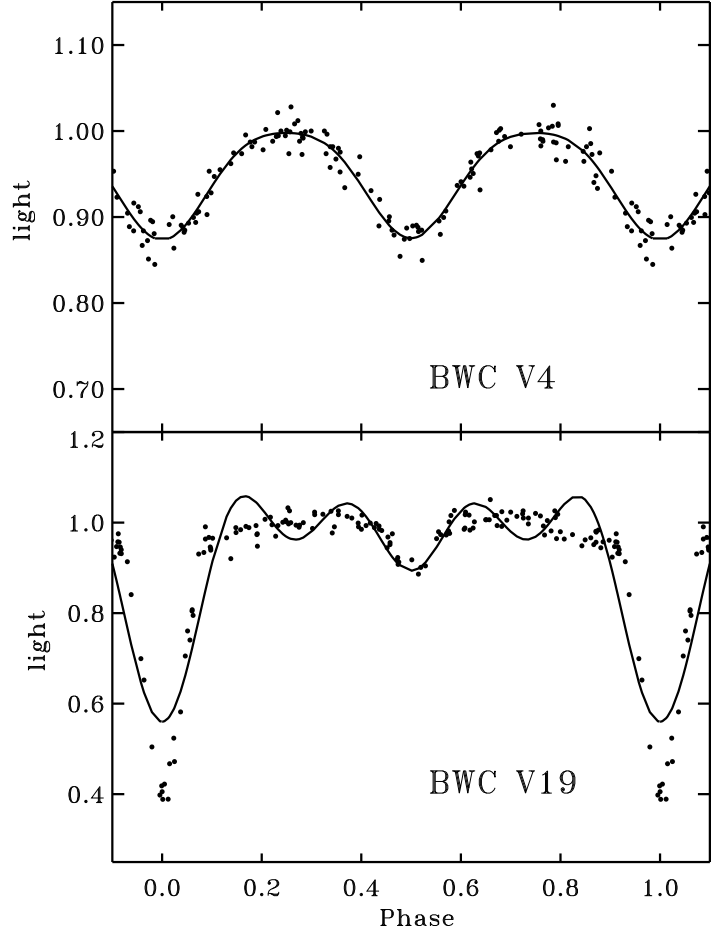,height=4.5in}}
\vskip 0.5in
\caption{Typical Fourier fits illustrated for the first EW and EA
systems in the OGLE sample, V4 and V19 in BWC (or in our numbering
scheme: \#0.004 and \#0.019). The even Fourier coefficients which are 
used in our automatic 
classifier are: $a_2 = -0.062$ and $-0.121$, and $a_4 = -0.011$ and
$-0.087$, respectively. These 
are fairly typical values for each class of objects. The odd
coefficients hold a potential for classification of light curves for
detached binaries. For \#0.019, $a_1 = -0.098$ and $a_3 = -0.069$,
whereas they are both equal to zero for the contact binary \#0.004.}
\end{figure}

\begin{figure}           
\centerline{\psfig{figure=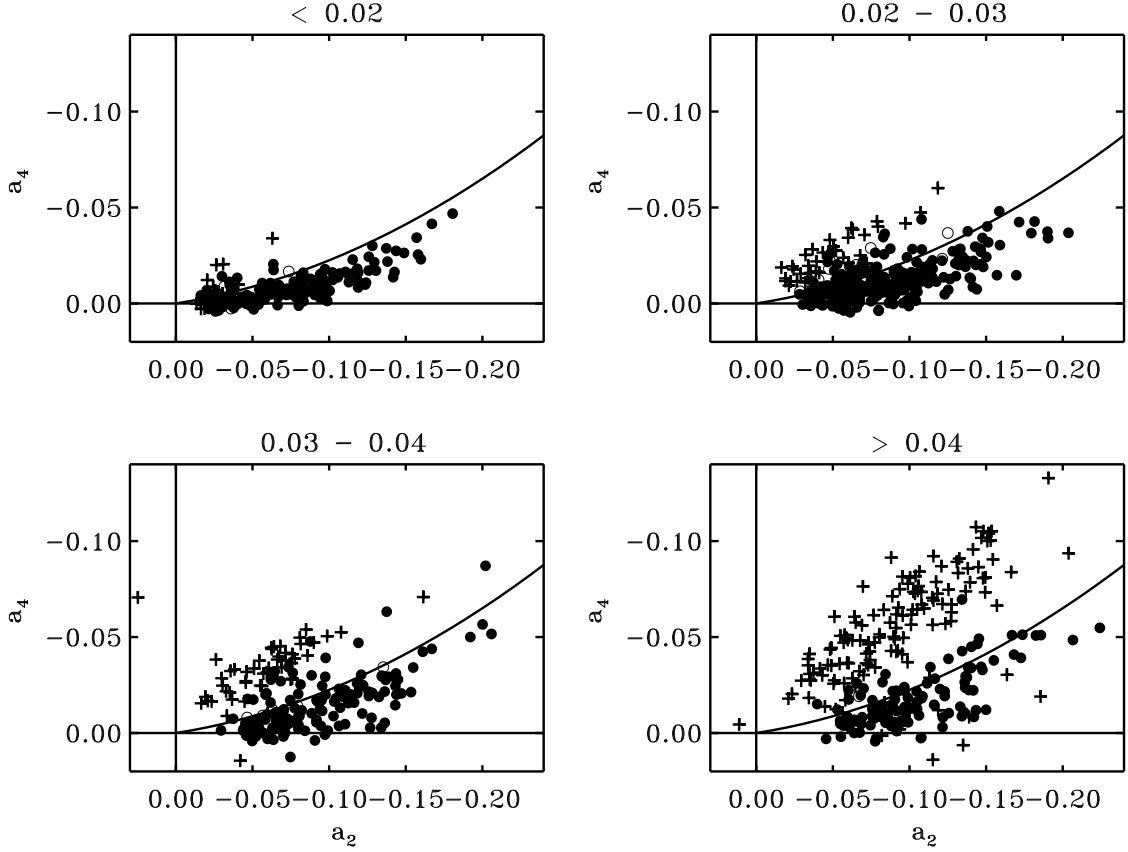,height=4.5in,angle=90}}
\vskip 0.5in
\caption{The automatic classifier uses the coefficients $a_2$ and
$a_4$ as well as the total quality of fit, measured by the mean
standard error per observation $\epsilon$
(ranges given in the panel headers). We selected contact
binaries by taking systems located below the marginal-contact curve,
with both $a_2$ and $a_4$ negative, and
with $\epsilon < 0.04$. Note, that all systems appearing in the lower-right
panel have been eliminated by the last requirement.
Thus, we rejected most of the detached systems and
some genuine contact systems, but with poor observations. The symbols
are the same as in  Figure~4.}
\end{figure}

\begin{figure}           
\centerline{\psfig{figure=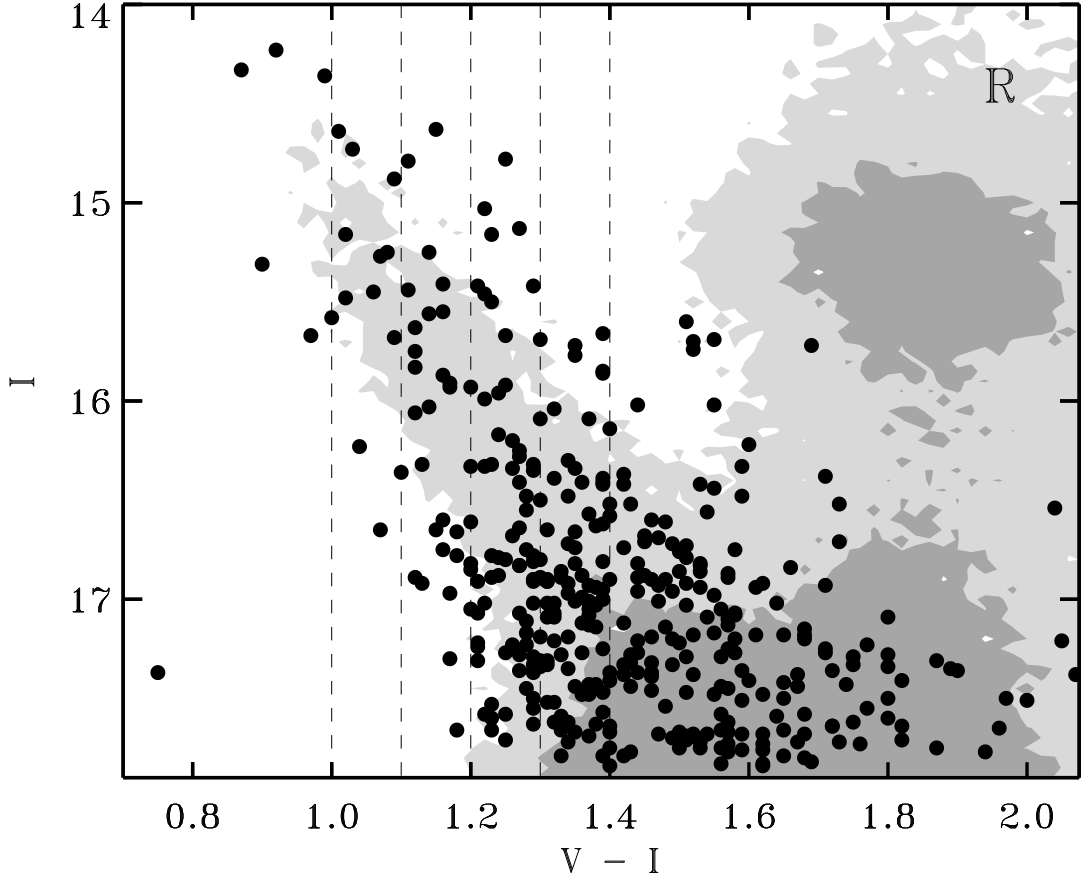,height=4.5in}}
\vskip 0.5in
\caption{The observed color -- magnitude diagram for the contact  
systems of the R-sample (filled circles), compared with the
distribution of the stars which were used to define the CMD for 
Baade's Window (two shades of grey for
10 and 40 stars per $(\Delta (V-I),\,\Delta I) = (0.02,\, 0.05)$). 
Majority of stars redder than $V-I \simeq 1.3 - 1.4$ 
(i.e. $(V-I)_0 \simeq 0.7 - 0.8$) 
are subgiants and giants of the Bulge. The Bulge Red-Clump
giants are particularly prominent at $V-I \simeq 1.8$ and 
$I \simeq 15.3$.  The Bulge Turn-Off Point is located at the lower
edge of the figure at $I \simeq 18$ (Kiraga et al. 1996). The slanted
sequence is due to Old Disk stars progressively reddened with distance
(see the text). The vertical dashed lines define bands 
for distributions of star numbers in Figure~8.}
\end{figure}

\begin{figure}           
\centerline{\psfig{figure=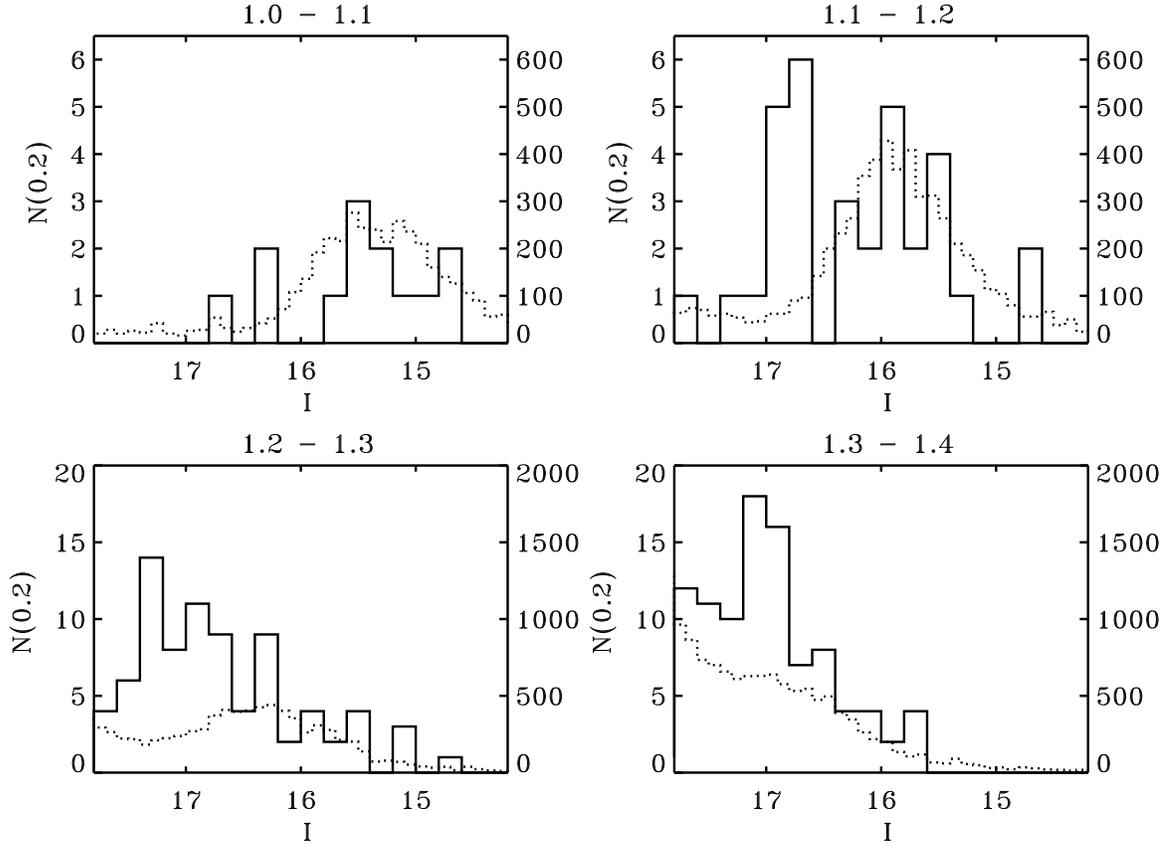,height=4.5in,angle=90}}
\vskip 0.5in
\caption{Distributions of the numbers of W~UMa-type systems as function
of observed $I$-magnitude in 4 vertical bands across the CMD, as
shown in Figure~7. The observed color ranges, within 
$1.0 < V-I < 1.4$, as given in headers for each of the panel.
The numbers are expressed per 0.2 mag in $I$ and shown as continuous
lines. The figure gives also the total numbers of all stars used in
Udalski et al. 1993 and P94 in analyses of the CMD's, also
converted to the same intervals, but sampled with two-times
smaller steps (dotted
line). The ratio of 100 for both groups of stars was selected quite
arbitrarily for convenience of plotting, but it does reflect the
approximate apparent frequency of W~UMa systems among ordinary stars
in the direction of the Bulge.
Statistical tests show that the {\em
shapes of the distributions\/} are basically
identical within each panel (see the text).}
\end{figure}


\begin{figure}           
\centerline{\psfig{figure=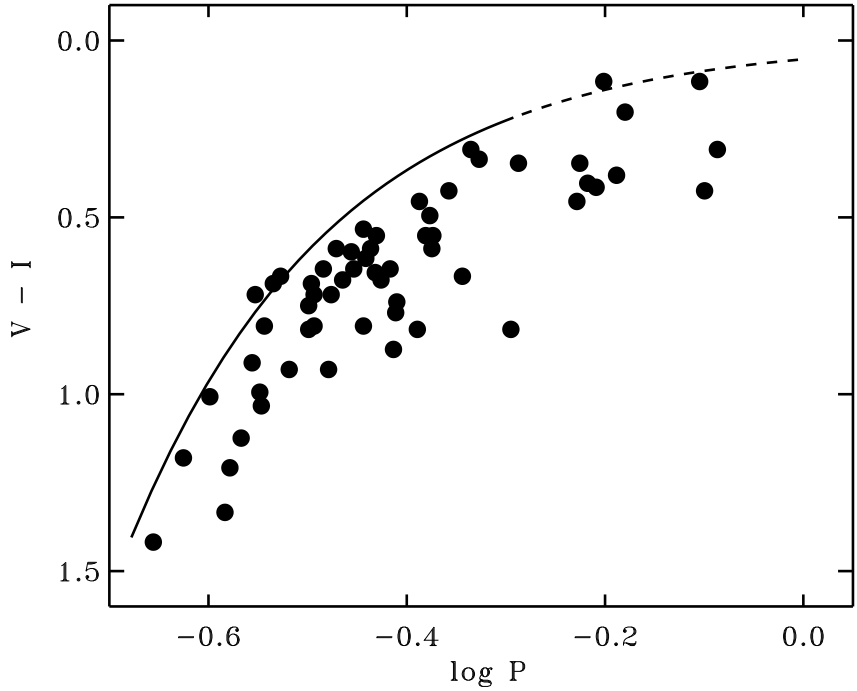,height=3.5in}}
\vskip 0.5in
\caption{The short-period/blue envelope (SPBE) in the period--color diagram 
has a special significance for our further considerations. It is shown
here, approximated by a simple expression, $V-I = 0.053 \,P^{-2.1}$,
with the period $P$ in days, together with nearby field systems from
the compilation of Mochnacki (1985). The de-reddened
$B-V$ colors given in this
compilation have been transformed to $V-I$ using the Main Sequence
relations of Bessell (1979, 1990), which we found adequately represent the
color--color relations for W~UMa-type systems. The shape of SPBE is
uncertain for $P > 0.5$ day. 
}
\end{figure}

\begin{figure}           
\centerline{\psfig{figure=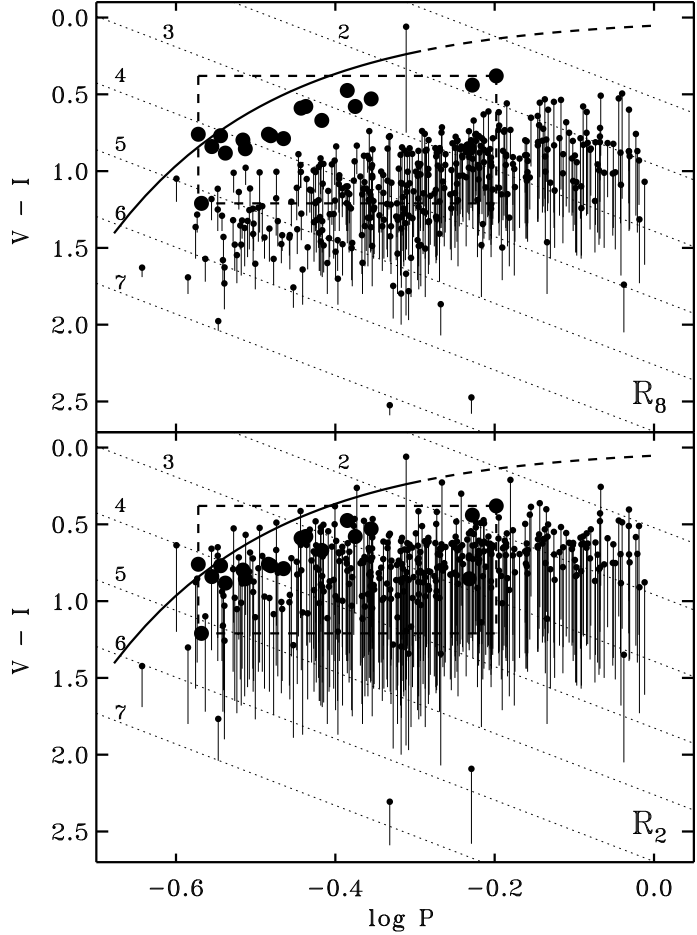,height=4.5in}}
\vskip 0.5in
\caption{The period -- color relation for systems of the R-sample for
two extreme assumptions on the extent of the dust layer of 8 kpc
(upper panel, called here R$_8$) and 2 kpc (lower panel, R$_2$). 
The observed and de-reddened values of $V-I$ 
are given by the lower and upper ends of the reddening vectors. 
Systems used in the most recent $M_I\,(\log P, V-I)$ calibration are
shown by large filled circles. They are located within a box (broken
line) where the calibration should be valid to better than 0.5
mag. The loci of equal $M_I$ according to that calibration are shown
by the inclined, dotted lines. 
The thick curve gives the SPBE, as described in previous figure. Note
that the gap between clustering of points and the SPBE for long-period
systems, which are, on the average, relatively faint
(cf.\ Figure 18), can be only slightly reduced by an assumption that
the OGLE data are affected by a systematic error in $V-I$ colors (see
Section 4.10.). 
}
\end{figure}

\begin{figure}           
\centerline{\psfig{figure=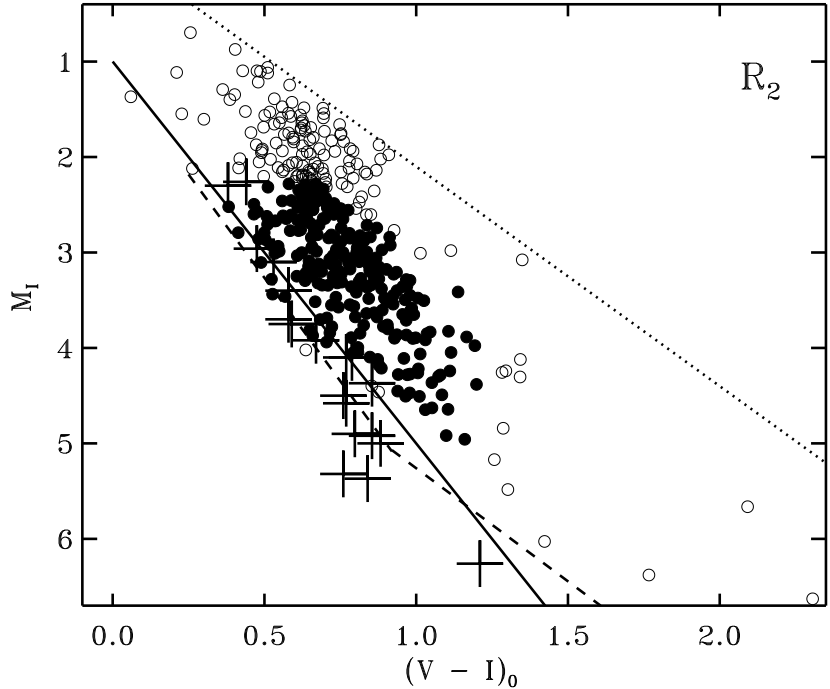,height=3.5in}}
\vskip 0.5in
\caption{The de-reddened color -- absolute magnitude diagram for contact
systems within (filled circles) and outside (open circles)
the range of the strict
applicability of the $M_I\,(\log P,\,V-I)$ calibration. Lines
give the Main Sequence relations for Pleiades (P94) (continuous)
and for nearby stars (Reid \& Majewski 1993) (broken). 
The dotted line gives the upper bound imposed by the 
1-day period limit which is conventionally used for W~UMa-type
systems. The calibrating systems are marked by large crosses. Note
that the slope for these systems is steeper than for single stars
reflecting contribution of the orbital period to changing geometry of 
the contact configuration. This systematic trend and part of 
the vertical
scatter are absorbed by the term $\propto \log P$ in the absolute
magnitude calibration. As we can see in this figure, 
the Baade's Window systems deviate systematically from the calibrating
systems, an effect which might be interpretted either as a genuine
tendency to
see mostly evolved systems at large distances, or by a presence of a
small systematic color error in the OGLE photometry. The latter
suspicion is discussed further in Section 4.10. Experiments with colors
artificially modified for $I > 15$ indicate a very small change in
this effect, mostly due to the complex nature of our scheme which
involves simultaneous determinations of absolute magnitudes, distances
and values of reddening. 
}
\end{figure}

\begin{figure}           
\centerline{\psfig{figure=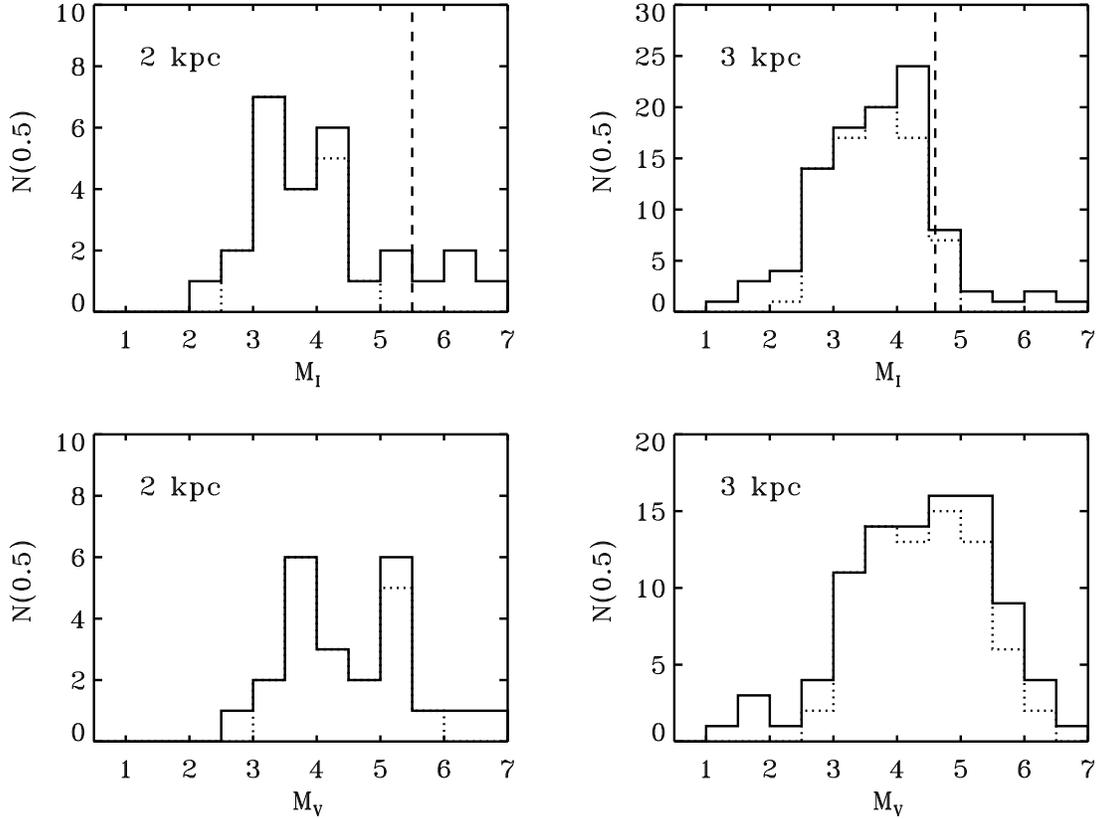,height=4.5in,angle=90}}
\vskip 0.5in
\caption{The luminosity function for systems of the R$_2$-sample
closer than 2 kpc (left panels) and 3 kpc (right panels). 
The dotted  lines mark the subsample of systems
falling within the ranges of periods, colors and absolute magnitudes
where the current $M_I$ calibration is applicable. The vertical broken
lines in the upper panels 
show the absolute magnitude limits $M_I$ of 5.5 and 4.6 for 
$I_{max} = 17.9$, $A_I = 0.9$.
}
\end{figure}

\begin{figure}           
\centerline{\psfig{figure=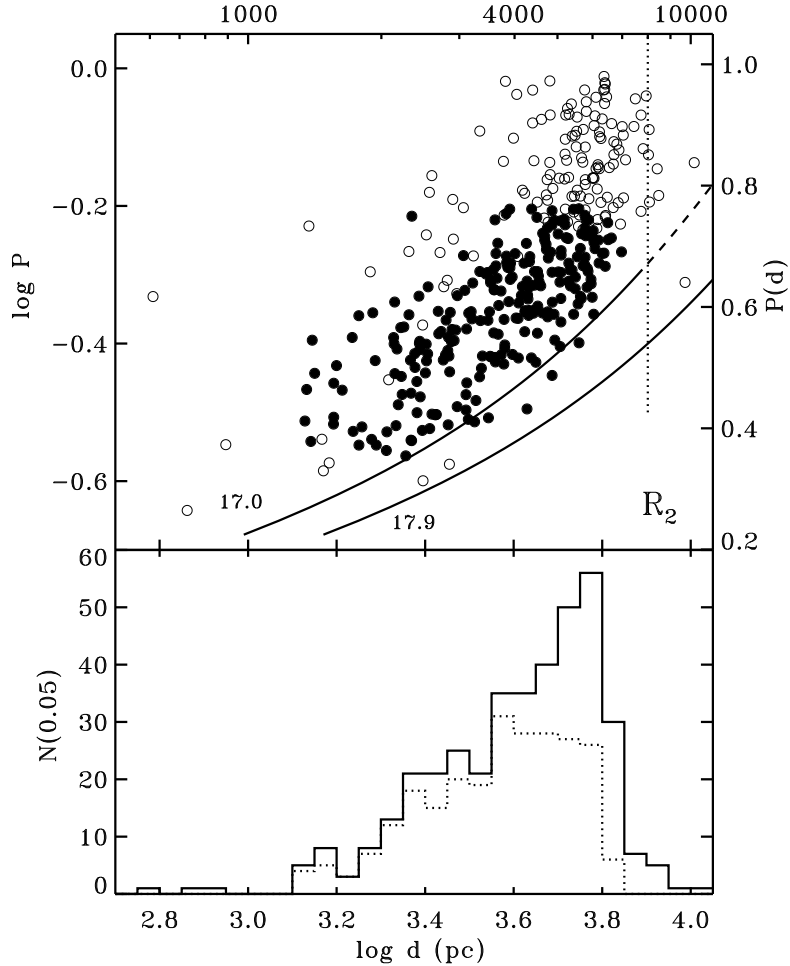,height=4.5in}}
\vskip 0.5in
\caption{Distribution of distances for the R$_2$-sample. The upper panel
shows a correlation of distances with orbital periods indicating very
clearly  that at large
distances only long-period systems remain visible. The crucial
importance of the SPBE on the period--color
relation in eliminating distant blue systems
is shown by two lines corresponding to the limiting magnitude
of $I=17.9$ and to the perceived level where some selection against
discovery might set in  at $I=17.0$. The filled circles
mark systems within the limits of periods, colors and absolute
magnitudes where the $M_I$ calibration is valid and good to better
than about 0.5 mag. The vertical dotted line gives the expected
location of the Bulge at 8 kpc. Note a hint of some deficiency between
6.5 and 8 kpc. The lower panel shows
the histograms for the full sample and for systems within the limits of
the calibration (dotted). We note that this diagram, as well as the
two following ones would be affected by presence of a color error
which is discussed in Section 4.10.: Experiments with artificially
modified data indicate that the systems would tend to spread
to larger distances, up to 10 kpc, and the concentration at 8 kpc as
well as the gap between 6.5 -- 8 kpc would disappear.}
\end{figure}

\begin{figure}           
\centerline{\psfig{figure=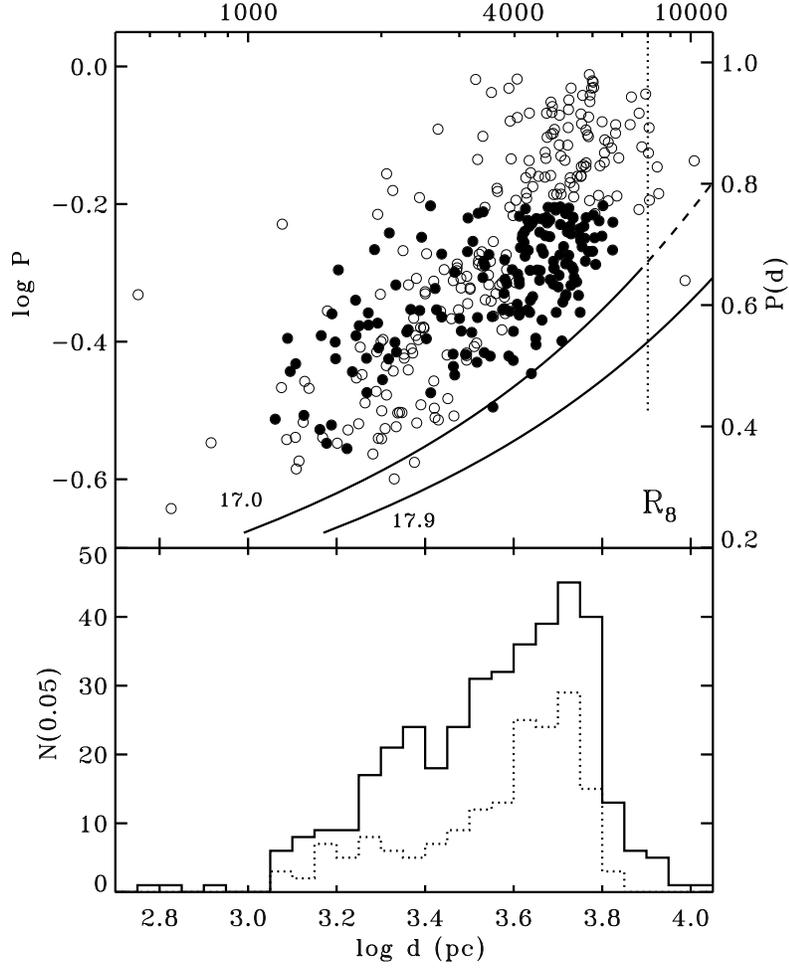,height=4.5in}}
\vskip 0.5in
\caption{Distribution of distances for the R$_8$-sample, in the same
format as for the R$_2$-sample in Figure~12. Note the flatter
distribution of distances than in R$_2$. 
Note also lack of any change in the gap
between 6.5 and 8 kpc which results from the fact that
the reddening corrections for these distant stars are practically
identical under both assumptions on the spatial distribution of
reddening.} 
\end{figure}

\begin{figure}           
\centerline{\psfig{figure=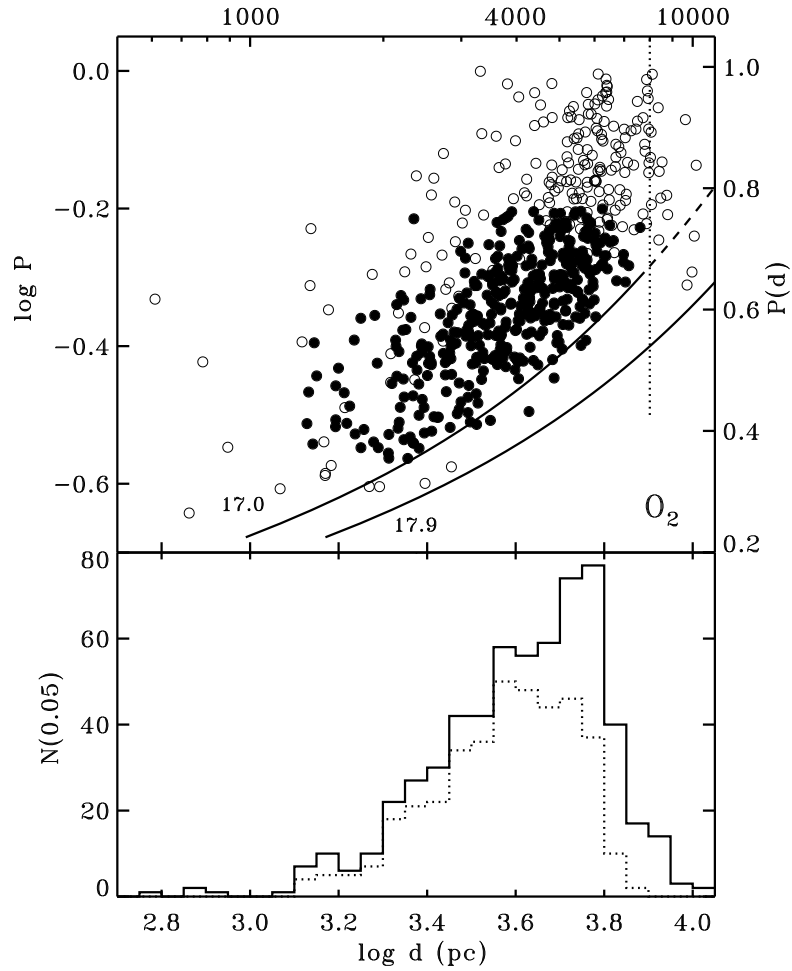,height=4.5in}}
\vskip 0.5in
\caption{Distribution of distances for the O$_2$-sample, in the same
format as for the R-sample in Figures~12 and 13. 
Apparently, both samples show
basically the same picture of uniform density of systems to distances
as large as 6 -- 7 kpc.}
\end{figure}

\begin{figure}           
\centerline{\psfig{figure=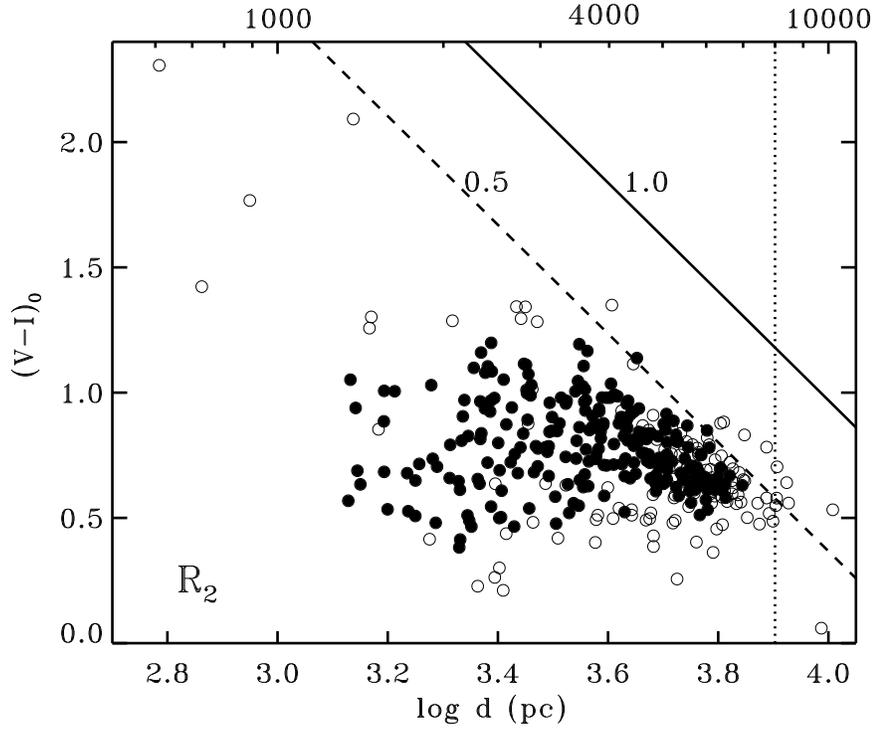,height=3.5in}}
\vskip 0.5in
\caption{Relation between the distance and the de-reddened color
$(V-I)_0$
for the R$_2$-sample. The slanted lines give the expected depth
limits for faint, red systems
at $I_{max} = 17.9$, for periods of 0.5 and 1 day, under an
assumption of uniform reddening and extinction, $E_{V-I} = 0.6$ and
$A_I = 0.9$. Note the concentration of colors in a relatively narrow
range $0.4 < (V-I)_0 < 1.0$, with the blue edge basically identical to
that of the Turn-Off Point of an old population,
and a relatively small number of very red systems.
}
\end{figure}

\begin{figure}           
\centerline{\psfig{figure=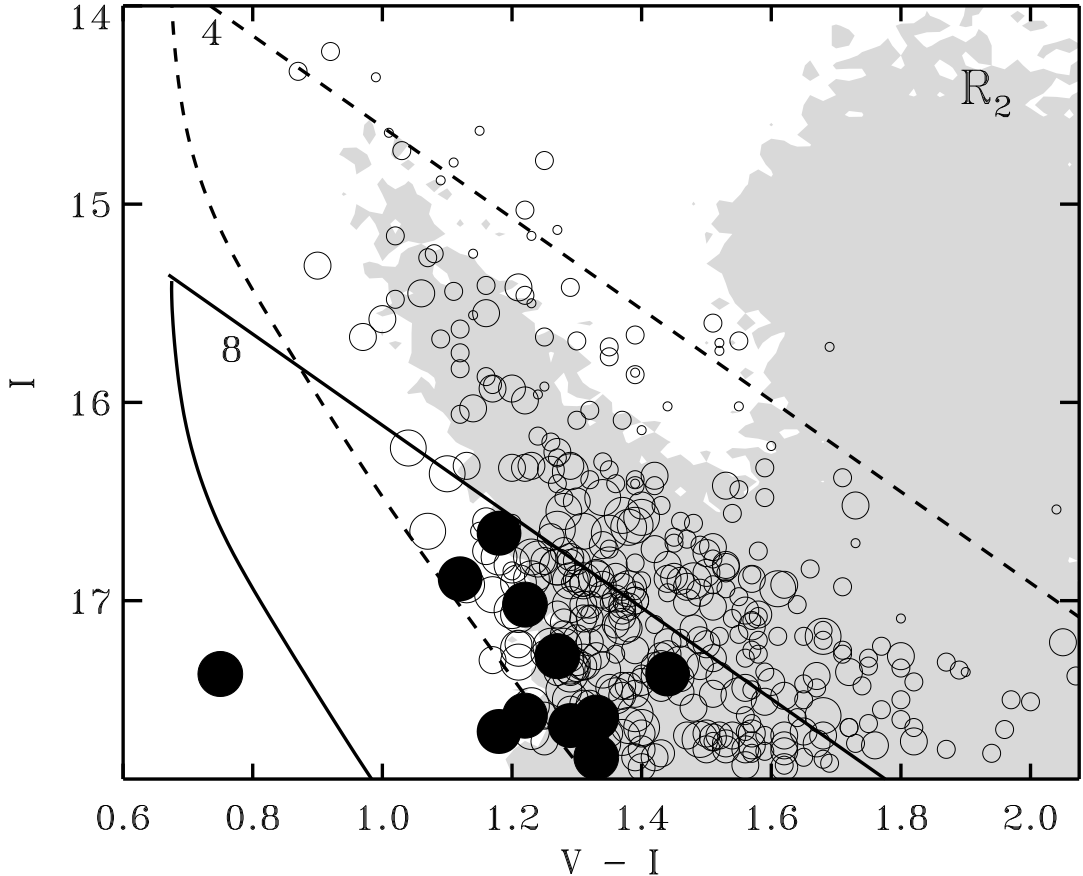,height=4.5in}}
\vskip 0.5in
\caption{The color--magnitude diagram similar to that in Figure~7 but
with distances of systems shown
by sizes of circles, incremented in 2 kpc intervals (the smallest
symbols are for $d < 2$ kpc).  Because the symbols heavily 
overlap, open circles have been used for all systems. The
exception are systems with $D > 7.5$ kpc which are shown by filled
circles. (The large distance to the blue system \#6.121 
might be entirely erroneous, cf.\ Sec.4.1.). To illustrate that no
selection effects should operate against discovery of distant systems
in the blue part of the diagram, we show two known limitations:
the arbitrary orbital-period limit of one day and
the impossibility of having blue/short-period systems beyond the SPBE
line. These two constraints result in existence of
``catchment areas'' which are shown for two distances of 4 and 8 kpc
as examples. The meaning of such an area
is as follows: For a given distance, to be included in the sample, the
system must be located below the straight line corresponding to the 1-day
period cutoff, but must be above the curved SPBE on 
the period--color relation. 
The grey region, as in Figure~7, corresponds
to more than 10 CMD-stars per color/magnitude interval.
}
\end{figure}

\begin{figure}           
\centerline{\psfig{figure=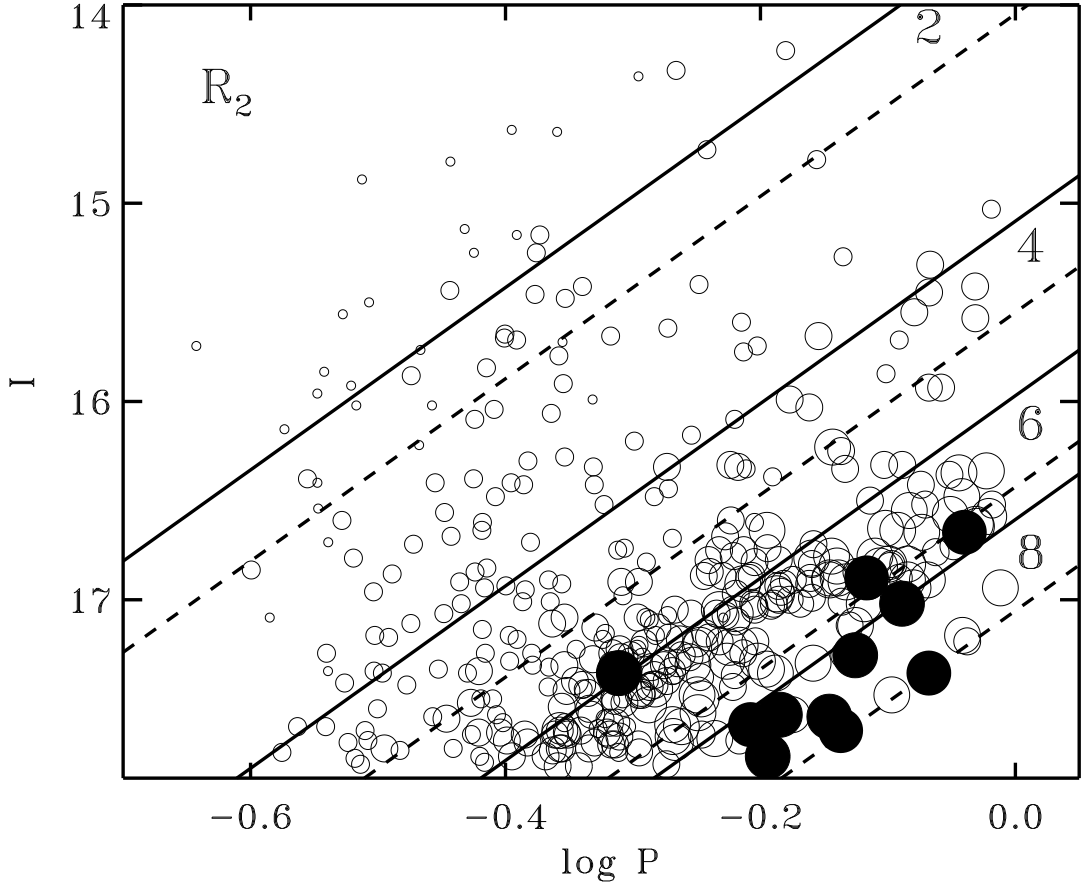,height=4.5in}}
\vskip 0.5in
\caption{Correlation of observed magnitudes with periods and
clustering in the lower-right corner of the period -- observed
magnitude diagram find an explanation in an increase in numbers 
of systems at large
distances. The lines are for distances of 2, 4, 6 and 8 kpc, and for
two values of $V-I$ of 1.2 (solid lines) and 1.4 (broken lines) with
the average values of $A_I = 0.9$ and $E_{V-I}=0.6$. The sizes of
symbols are as in Figure 17.
}
\end{figure}

\clearpage    

\begin{figure}           

\centerline{\psfig{figure=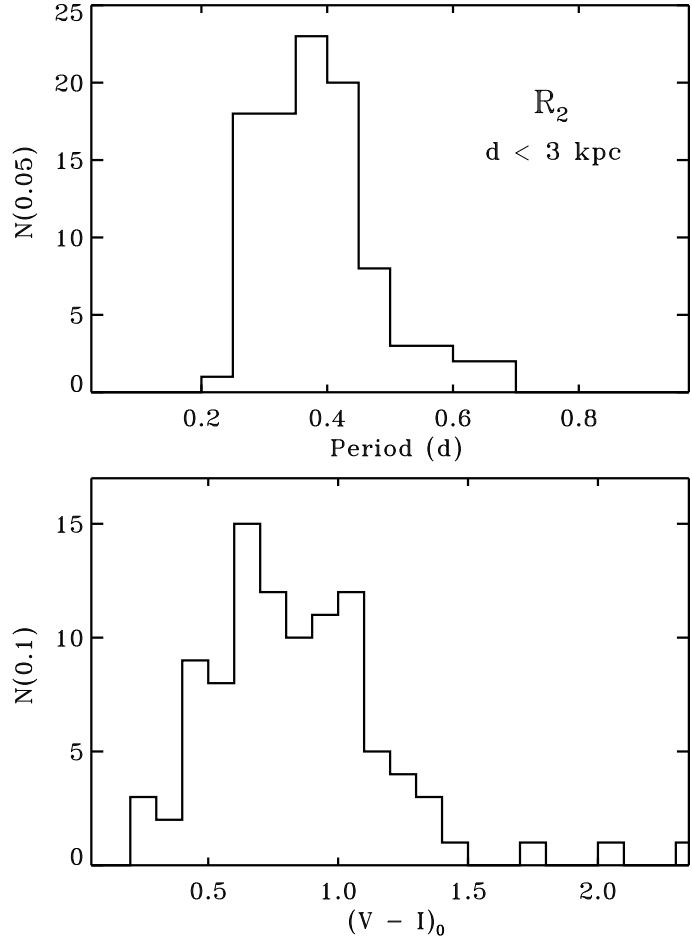,height=4.5in}}
\vskip 0.5in
\caption{Distributions of orbital periods and intrinsic colors for the
sample R$_2$, limited to the distance of 3 kpc. This sample should be
complete and free of distance-related biases.}

\end{figure}

\end{document}